\documentclass[conference]{IEEEtran}
\IEEEoverridecommandlockouts
\usepackage{cite}
\usepackage{amsmath,amssymb,amsfonts}
\usepackage{algorithmic}
\usepackage{graphicx}
\usepackage{textcomp}
\usepackage{xcolor}
\usepackage{upgreek}

\usepackage{multirow}
\usepackage{enumitem}
\usepackage{subcaption} 

\usepackage{booktabs}
\usepackage{multirow}

\usepackage{xspace}
\usepackage[compatibility=false]{caption}

\def\BibTeX{{\rm B\kern-.05em{\sc i\kern-.025em b}\kern-.08em
    T\kern-.1667em\lower.7ex\hbox{E}\kern-.125emX}}
\begin{document}

\title{Adversarial Pre-Padding: Generating Evasive Network Traffic Against Transformer-Based Classifiers}


\author{
\IEEEauthorblockN{
Quanliang Jing$^{1}$,
Xinxin Fan$^{1,2}$,
Yanyan Liu$^{1}$,
Jingping Bi$^{1,2}$
}
\IEEEauthorblockA{
$^{1}$\textit{Institute of Computing Technology, Chinese Academy of Sciences, Beijing, China}\\
$^{2}$\textit{University of Chinese Academy of Sciences, China}\\
Email: \{jingquanliang, fanxinxin, yanyanliu, bjp\}@ict.ac.cn
}
}

\maketitle

\newcommand{\our}{\textsc{AdvTraffic }\xspace}

\newcommand{\ourNoSpace}{\textsc{AdvTraffic}}

\newcommand{\red}[1]{\textcolor{red}{\emph{\{#1\}}}}

\newcommand{\blue}[1]{\textcolor{blue}{\emph{\{#1\}}}}
\newcommand{\yellow}[1]{\textcolor{yellow}{\emph{\{#1\}}}}


\begin{abstract}
To date, traffic obfuscation techniques have been widely adopted to protect network data privacy and security by obscuring the true patterns of traffic. Nevertheless, as the pre-trained models emerge, especially transformer-based classifiers, existing traffic obfuscation methods become increasingly vulnerable, as witnessed by current studies reporting the traffic classification accuracy up to 99\% or higher. To counter such high-performance transformer-based classification models, we in this paper propose a novel and effective \underline{adv}ersarial \underline{traffic}-generating approach (AdvTraffic\footnote{The code and data are available at: https://anonymous.4open.science/r/TrafficD-C461/}). Our approach has two key innovations: (i) a pre-padding strategy is proposed to modify packets, which effectively overcomes the limitations of existing research against transformer-based models for network traffic classification; and (ii) a reinforcement learning model is employed to optimize network traffic perturbations, aiming to maximize adversarial effectiveness against transformer-based classification models. To the best of our knowledge, this is the first attempt to apply adversarial perturbation techniques to defend against transformer-based traffic classifiers. Furthermore, our method can be easily deployed into practical network environments. Finally, multi-faceted experiments are conducted across several real-world datasets, and the experimental results demonstrate that our proposed method can effectively undermine transformer-based classifiers, significantly reducing classification accuracy from 99\% to as low as 25.68\%.

\end{abstract}


\section{Introduction}\label{sec:Introduction}


Network traffic classification (NTC), serving as a fundamental method for traffic analysis, is frequently exploited by attackers for malicious activities \cite{8713803, bagui2017comparison}, such as launching network attacks, bypassing security measures, and conducting user-behavior assessments. To counter traffic identification, various obfuscation techniques have been developed to obscure traffic patterns by modifying packet size, direction, and time intervals. Existing research \cite{meier2022ditto,nasr2021defeating,rahman2020mockingbird,mohajeri2012skypemorph,DBLP:conf/sp/ShenJWLKXZ24} demonstrates that such traffic modifications can significantly degrade the performance of traditional feature-based classification methods. Nevertheless, recent advances in deep learning, particularly with transformer-based pre-trained models, have revolutionized network traffic classification by leveraging large and complex neural networks \cite{lin2022bert, LIN2024103892, s21248231, 10575448}. These models leverage large, unlabeled traffic datasets to learn unbiased traffic representations, then perform fine-tuning using a small amount of labeled data, by which the learned representations can be easily transferred to different downstream tasks like IP-packet and TCP-stream classification. For instance, the pre-trained model ET-BERT \cite{lin2022bert} demonstrated the ability to outperform all previous works, achieving 99\% accuracy for encrypted traffic (e.g., VPN/Tor). These models classify traffic based on byte sequences rather than traffic features, making changes in traffic patterns alone inadequate for defending against attackers exploiting these advanced classifiers.

As is well known, while pre-trained models offer significant advantages in learning latent features from various data modalities, such as images, text, audio, and video, they remain susceptible to adversarial examples, which are produced by carefully-crafted perturbations on the clean data with the purpose of deceiving target models \cite{li-etal-2020-bert-attack, ren2019generating}. To date, distinguishing such subtly perturbed adversarial examples from clean counterparts remains challenging. This raises a natural question: \textbf{Can we use the adversarial examples\footnote{This study adheres to the attack–defense paradigm established in prior literature, wherein the classifier is regarded as the attacker and the proposed method serves as the defender. Traffic perturbed by the defender is accordingly denoted as adversarial examples (traffic).} of network traffic to resist the transformer-based pre-trained models to classify traffic?}

Adversarial examples are still an active and prominent research topic nowadays in various domains, including image recognition/detection and natural language processing (NLP), etc. However, due to the substantial differences between network traffic and other data modalities, applying adversarial techniques to the network traffic domain present three severe challenges: i) \textbf{Protocol-Rule Constraints.} Network traffic (i.e., packets) follows a rigid organizational structure governed by protocol specifications, such as those defined in the TCP/IP stack. Unlike adversarial perturbations in domains such as image or text, where elements can often be freely modified, alterations to network packets must adhere strictly to protocol constraints to maintain functional integrity. Any perturbation that violates these constraints risks disrupting packet delivery across intermediate network devices such as routers and switches.  To address this challenge, we design adversarial perturbations by selectively modifying protocol-compliant fields that do not interfere with the normal transmission process. This approach ensures that the adversarial effect is achieved without compromising the packet’s deliverability or violating protocol semantics; ii) \textbf{Traffic-Perturbation Strategy.} Existing traffic obfuscation mechanisms  typically perturb traffic by appending byte sequences at the end of packets. However, such post-padding strategies have demonstrated limited effectiveness against transformer-based pre-trained deep learning models, which are capable of learning robust representations despite these superficial modifications. To address this vulnerability, we propose a pre-padding strategy that introduces byte sequences at the beginning of packets. This approach allows for more meaningful perturbations that interfere with the semantic understanding of the traffic by pre-trained classifiers, thereby significantly degrading their classification performance; and iii) \textbf{Adversarial Perturbations Generating Manner.} Existing methods typically generate perturbations in a random manner, which presents significant limitations. Such randomly generated byte sequences are often suboptimal and lack systematic optimization, resulting in limited ability to degrade model performance effectively. In contrast, fields such as image recognition benefit from gradient-based methods for constructing effective adversarial examples. However, in the context of network traffic, this approach is infeasible due to the extensive preprocessing required to convert raw traffic into a format suitable for deep learning models, which breaks the differentiability chain and prevents gradients from being propagated back to the original byte sequences. To address this challenge, we propose a reinforcement learning-based strategy for generating adversarial examples. This method formulates the byte sequence generation process as a reinforcement learning task, enabling the model to explore the action space and generate targeted perturbations via byte-padding, guided by feedback from the network environment.

To surround the challenges, three main contributions are summarized as:
\begin{itemize}
	\item 
        We are the first to analyze the root causes why the conventional defense methods fail to defend against the transformer-based pre-trained NTC models, and upon which we propose a pre-padding strategy to generate effective adversarial perturbations on network traffic to defeat such pre-trained classifiers.
        
	\item 
        Adversarial traffic is generated by selectively modifying specific fields in a manner that preserves normal network communication, while reinforcement learning is employed to maximize resistance against transformer-based pre-trained NTC models.
        

        
	\item
        Extensive experiments are conducted on three real-world datasets. The experimental results demonstrate our proposed \our significantly outperforms the state-of-the-art baselines and remarkably reduces the classification accuracy of transformer-based pre-trained NTC models from over 99\% to below 25.68\%.
    

   
       

\end{itemize}


\section{Problem Statement}\label{sec:threadDefenseModel}

A network flow can be formally represented as an ordered sequence of packets $P=[P_1, P_2, ..., P_n]$. Where $P_i$ denotes an $i$-th packet, and $n$ represents the total number of packets in the flow. 


\textbf{Our Problem: Defeating Transformer-Based Pre-Trained Traffic Classifiers.} Deep learning models are vulnerable to adversarial examples that introduce subtle adversarial perturbations to mislead predictions. Unlike prior methods that primarily target traffic patterns \cite{shenoi2023ipet,zolbayar2022generating,rahman2020mockingbird,ling2022towards}, this paper investigates the feasibility of defending transformer-based pre-trained classifiers by perturbing packet-level byte sequences instead of flow-level statistical features. Let $f(\cdot)$ denote a classifier that maps an input packet $\mathbf{x}$ to a label $y \in \Phi$, where $\Phi$ is the set of classification labels. Adversarial perturbations can be formalized as the following optimization problem:

\begin{equation} \label {equation-00}
    \mathbf{x^*} = \mathbf{x} + argmin\{\mathbf{\delta}: f(\mathbf{x+\delta}) \ne f(\mathbf{x})\} = \mathbf{x} + \mathbf{\delta}_x,
\end{equation}
where $\mathbf{\delta}_x$ is the perturbation added to $\mathbf{x}$, and $\| \mathbf{\delta}_x \|$ denotes the norm of the perturbation, representing the magnitude of change. The inequality $f(\mathbf{x+\delta}) \ne f(\mathbf{x})$ presents that the perturbed input $\mathbf{x} + \mathbf{\delta}_x$ is misclassified by the classifier.

\textbf{Does The Perturbation Amplitude Need To Be Constrained?} For network traffic, modification amplitude refers to the number of bytes modified relative to the original byte sequence. It is desirable to maximize the adversarial effectiveness while reducing modification amplitude to avoid excessive traffic, which may degrade network performance and user experience.




\subsection{\textbf{Threat Model}}

Our adversarial perturbation technique is specifically tailored to defend against Transformer-based traffic classification models, with particular emphasis on ET-BERT\cite{lin2022bert}, which achieves state-of-the-art accuracy by modeling packet byte sequence representations. Beyond ET-BERT, our method also demonstrates robustness against other Transformer-based or Transformer-inspired classifiers, such as NetMamba\cite{DBLP:conf/icnp/WangXWWZ024} and YaTC\cite{DBLP:conf/aaai/0001ZDWWGX23}, underscoring its general applicability to a broader family of deep sequence models.




\subsection{\textbf{Defense Model}}\label{DefenseModel}


\textbf{Defender’s knowledge of the target traffic.} We assume that the defender lacks prior knowledge regarding the content or traffic patterns of the target network packets subjected to perturbation.


\textbf{Defender’s knowledge of the model.} We consider two defense scenarios: white-box and black-box settings.

\begin{enumerate}[label={\alph*)}]

   \item \textbf{White-Box Defense.} The defender has full knowledge of the NTC model architecture and associated parameters used by the attacker.

   \item \textbf{Black-Box Defense.} The defender has access only to the inference labels of the NTC model, without knowledge of its internal architecture, parameters, or gradients.
   
\end{enumerate}


\textbf{The defender’s goal.} In this study, we focus exclusively on untargeted defense, wherein the defender perturbs clean traffic to induce the NTC model to misclassify it into any incorrect category.



 




\section{The Proposed Model}\label{sec:Model}


\begin{figure*}[ht]
	\centering
	\includegraphics[scale=0.52]{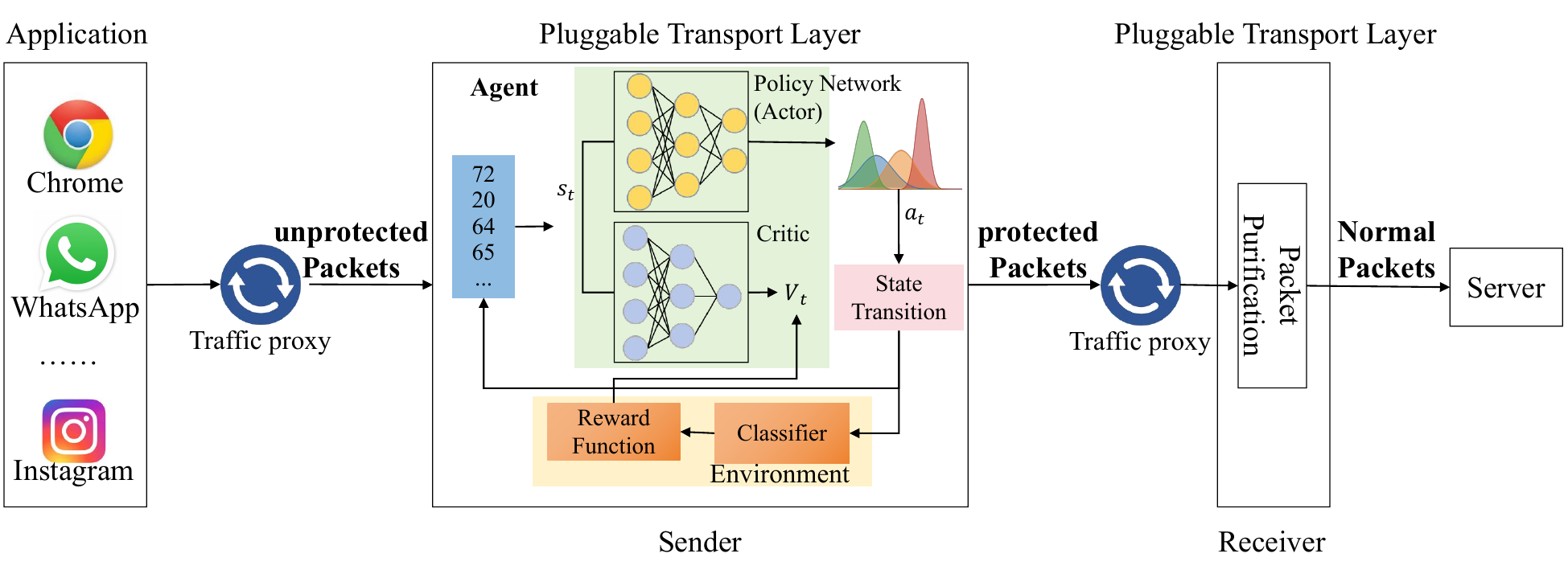}  
	\caption{The framework of \ourNoSpace. \our applies adversarial perturbations to network traffic by utilizing an agent to alter semantics. These perturbations are designed to deceive traffic classifiers while maintaining communication efficiency (The figure illustrates the client-to-server operations, with server-to-client operations following a similar process).}
	\label{fig:framework}
\end{figure*}



\subsection{\textbf{Framework Overview}}\label{sec:overview}

As illustrated in Figure~\ref{fig:framework}, the proposed framework comprises two core components, Sender and Receiver, which are implemented as pluggable network-layer modules. The Sender perturbs traffic from applications (e.g., Chrome, Instagram) by modifying IP packets, while preserving network protocols (e.g., TCP/IP) and ensuring semantic integrity. The Receiver reverses these perturbations to restore packets for correct parsing and delivery to upstream applications. This de-perturbation process relies on shared knowledge of modified packet indices, achievable via dedicated protocol design.


\subsection{\textbf{Vulnerability Analytics for Existing Adversarial Perturbation Methods}}\label{sec:existingMethods}

\begin{figure*}[hbt]
	\centering
	\includegraphics[scale=0.45]{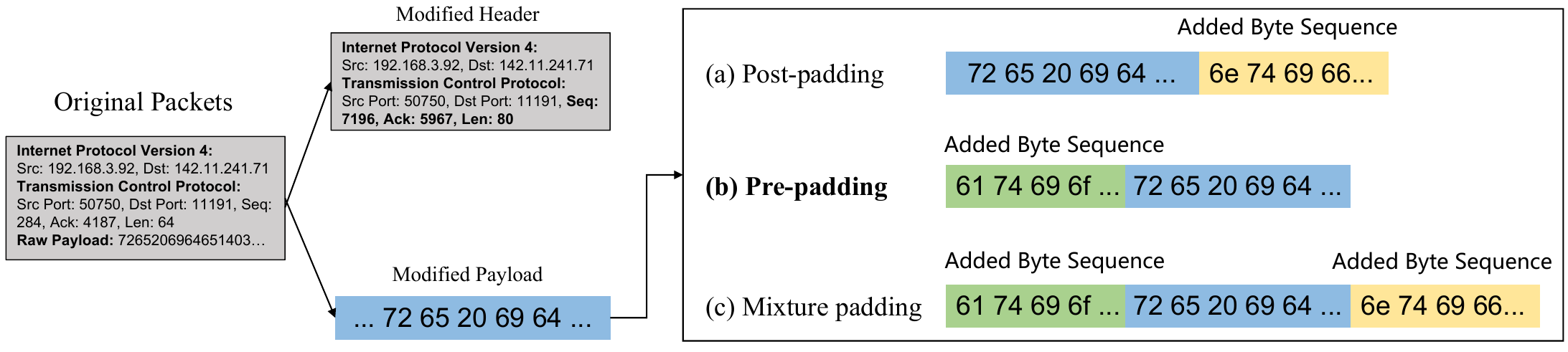}
	\caption{Illustration of Traffic Padding Strategies. To maintain protocol compliance, the IP packet's length and checksum fields are dynamically recomputed based on the modified packet.}
	\label{fig:paddingStyle}
\end{figure*}

\textbf{Applying Existing Methods in Semantics Defense.}
Various adversarial techniques have been developed to obfuscate network traffic and evade classification. Representative methods include Mockingbird \cite{rahman2020mockingbird}, BLANKET \cite{nasr2021defeating}, Walkie-Talkie \cite{wang2017walkie}, and padding-based defenses such as iPET \cite{shenoi2023ipet} and DeTorrent \cite{holland2023detorrent}. These approaches predominantly manipulate the statistical features of traffic to evade DNN-based traffic classifiers. Our analysis reveals that existing packet perturbation techniques largely depend on post-padding strategies, wherein adversarial byte sequences are appended to the end of packets, as illustrated in Figure~\ref{fig:paddingStyle}(a). For example, BLANKET \cite{nasr2021defeating} adopts this technique by padding packets with random bytes to degrade the performance of classifiers.

To evaluate the impact of post-padding on degrading the performance of the transformer-based pre-trained traffic classifier ET-BERT~\cite{lin2022bert}, adversarial examples generated through random post-padding (Figure~\ref{fig:paddingStyle}(a)) are evaluated against the model trained on clean traffic. Experimental results indicate that this approach exerts minimal influence, with ET-BERT~\cite{lin2022bert} maintaining an accuracy of approximately 98\%. The findings underscore the limited effectiveness of naive padding strategies in misleading transformer-based pre-trained traffic classifiers.



This investigation yields a key insight: existing packet perturbation methods are largely ineffective against the pre-trained NTC model ET-BERT. These approaches fail to meaningfully alter the semantic representation of packets, thereby falling short in deceiving robust classifiers. Motivated by this limitation, we focus on proposing a novel adversarial perturbation technique specifically designed to target the transformer-based pre-trained NTC model, ET-BERT.


\textbf{Why existing methods are ineffective?} To dig out the behind reason why the current adversarial perturbation methods are ineffective, we conduct a thorough study on the underlying principles of the representative transformer-based pre-trained NTC model ET-BERT. Our findings unveil  the ET-BERT’s classification relies predominantly on the initial byte sequences of a network packet, while the later byte sequences have minimal impact on the classification outcome.


\begin{table*}[]
	\centering
	\caption{Classification Accuracy at Different Input Lengths}	\label{accuracyfordifferentinputlengths}
    \setlength{\tabcolsep}{2.5mm}
\begin{tabular}{|l|c|c|c|c|c|c|c|c|}
\hline
Input Length       & 1      & 2      & 4      & 8      & 16     & 32     & 64     & 128    \\ \hline
CSTNET-TLS         & 0.0088 & 0.3801 & 0.4945 & 0.8718 & 0.9267 & 0.9995 & 0.9995 & 0.9995 \\ \hline
ISCX-VPN           & 0.0833 & 0.3483 & 0.3497 & 0.6077 & 0.8363 & 0.9942 & 0.9963 & 0.9968 \\ \hline
\end{tabular}
\end{table*}

To validate this hypothesis, we conduct experiments by truncating network packets and inputting only the first $N$ bytes into the classification model. As shown in Table~\ref{accuracyfordifferentinputlengths}, classification accuracy improves with longer input lengths but plateaus at 32 bytes, which suggests that the first 32 bytes contain sufficient discriminative features for accurate classification. Consequently, post-padding beyond this region has negligible impact on model predictions, explaining the ineffectiveness of such perturbation techniques.


\textbf{New strategies for modifying packet semantics.} Unlike prior methods, our approach, illustrated in Figure~\ref{fig:paddingStyle}, introduces adversarial byte sequences at two strategically selected positions: within the transport layer header and in the boundary region between the header and payload, as shown in Figure~\ref{fig:paddingStyle}(b), rather than at the packet's end. This design is motivated by the observation that transformer-based pre-trained NTC models primarily extract features from initial byte sequences. By perturbing this semantically critical region, our strategy effectively disrupts the model’s learned representations while preserving protocol compliance, thereby improving adversarial effectiveness. Besides, as shown in Figure~\ref{fig:paddingStyle}(c), the proposed method is compatible with existing defense methods and defends against both statistical and pre-trained classifiers.

Specifically, for TCP packets, perturbations are applied to non-critical header fields—sequence number, acknowledgment number, window size, and urgent pointer—which do not affect transmission, as IP-layer forwarding relies solely on source/destination addresses and the IP checksum. To enable correct reconstruction at the server, original values of the modified fields are appended to the packet's end. In contrast, UDP headers remain unmodified due to their minimal structure. Finally, for both TCP and UDP, adversarial byte sequences are inserted immediately before the payload. All dependent fields, including length and checksums in the IP, TCP, and UDP headers, are recalculated to ensure protocol compliance.


\subsection{\textbf{Adversarial Packet Semantics Generation with Reinforcement Learning}}\label{rlAdv}


Although our pre-padding strategy proves highly effective, it also presents a significant challenge: how to select adversarial byte sequences that maximize the adversarial effect. Previous approaches typically relied on randomly generated byte sequences, which fail to guarantee optimal adversarial performance. To address this challenge, we propose a deep learning-based approach that generates highly adversarial byte sequences tailored to individual packets. Concretely, under the untargeted setting the generator seeks to induce misclassification into any incorrect class, thereby degrading the classifier’s overall performance across multiple categories.



Figure~\ref{fig:framework} illustrates the proposed method, which trains a deep reinforcement learning (RL) agent to iteratively perturb network packets. We formulate adversarial packet generation as a sequence generation task, where the agent receives a byte sequence at each timestep and outputs a corresponding perturbation action. This approach enables on-the-fly generation of adversarial packets without requiring full flow collection. Importantly, \our preserves protocol functionality, leaving mechanisms such as handshakes, error handling, and acknowledgments untouched. The method modifies only the semantic content—via header and payload changes—ensuring the resulting flow remains compliant with TCP/UDP specifications.


Concretely, given a packet $P_i$, the defender’s process $\tau$ is modeled as a Markov Decision Process (MDP) $M = (S, A, P, R)$, where $S = \{s_t\}$ denotes the state space, $A = \{a_t\}$ the action space, $P(s_{t+1} \mid s_t, a_t)$ the state transition function, and $R(s_t, a_t)$ the reward function. The defense episode is represented as $\tau = \{s_1, a_1, r_1, \dots, s_n, a_n, r_n\}$, with the initial state $s_1 = P_i$. The central challenge is to learn an optimal action $a_t$ for each state $s_t$. While the optimal policy could be framed as $P(a_t \mid s_t, \dots, s_1)$, this formulation incurs high computational cost due to long-term dependencies. To reduce complexity, we assume the Markov property: $P(a_t \mid s_t, \dots, s_1) = P(a_t \mid s_t)$, treating each packet state independently. Under this simplification, reinforcement learning (RL) is used to train a policy network that maps states to actions, optimizing for adversarial effectiveness. The core RL components are defined as follows:

\textbf{State Space.} It includes all intermediate states arising from valid perturbations, encompassing the initial victim packet, transitional states during defense, and the final perturbed state.


\textbf{Action Space.} The action space comprises all valid byte-level insertions within network packets. Accordingly, each time step offers 256 discrete actions, each corresponding to the insertion of a distinct byte value.


\textbf{State Transition Function.} Given an action $a_t$ in state $s_t$, the next state $s_{t+1}$ is generated by either modifying selected header fields or inserting the byte corresponding to $a_t$ immediately before the payload.


\textbf{Reward Design.} In this study, we investigate two distinct defense scenarios—white-box and black-box settings—and design tailored reward functions for each to optimize the adversarial perturbation process accordingly.

In the white-box defense scenario, we assume access to the model’s output probability distribution for a given state $s_t$ (e.g., via substitute models \cite{DBLP:conf/ccs/PapernotMGJCS17,Zhang_2022_CVPR}, though their construction is beyond this paper’s scope). The defender aims to alter $f(\cdot)$ such that $f(s_{t+1}) \neq f(s_t)$. Accordingly, the reward function is defined based on the divergence between predicted distributions, with greater disparity indicating higher misclassification likelihood. Additionally, the representation distance between $s_t$ and $s_{t+1}$ is incorporated to capture the perturbation's impact on classification. The overall reward $R(s_t, a_t)$ is thus defined as:


\begin{equation}\label {equation-15}
    R(s_t, a_t)=KL(f_{-1}(s_t),f_{-1}(s_{t+1})) + D(f_{-1}(s_t),f_{-1}(s_{t+1})),
\end{equation}
where $f_{-1}$ represents the probability distribution over the model's output at state $s_t$.


In the black-box defense scenario, the defender only observes the predicted label of the classifier for a given state $s_t$, with no access to internal model information. A positive reward is assigned if the attack causes misclassification; otherwise, the reward is zero. The reward function is defined as:

\begin{equation}\label {equation-15}
R(s_t, a_t)=\left\{
\begin{aligned}
1,\quad &  f(s_t) \neq f(s_{t+1})\\
0,\quad &  f(s_t) = f(s_{t+1})
\end{aligned}.
\right.
\end{equation}

\subsection{\textbf{Policy Network}}

The policy network (Actor) learns the policy $p(a_t|s_t)$ from the packet representation. Since the state at timestep $t$ incorporates all perturbations applied from steps 1 to $t-1$, its length grows over time, making fixed-size-input models like MLPs unsuitable. To handle variable-length states, we employ a BERT-based encoder leveraging self-attention to produce a fixed-size embedding $E_t$ from arbitrarily long byte sequences. The action probability distribution is then generated from $E_t$, and actions are sampled accordingly:


\begin{equation}\label {equation-16}
D_t=Softmax(\frac{\mathrm{MLP}(E_t)}{\tau}),
\end{equation}

\begin{equation}\label {equation-16}
a= Sample(D_t),
\end{equation}
where $Softmax$ is a normalization function that scales the probability of each action to an interval $[0, 1]$. Here $Sample$ refers to the action-sampling function, and $a$ denotes the specific action taken.

We incorporate a temperature coefficient $\tau$ into the Softmax function to regulate action selection diversity. When $\tau < 1$, the output distribution sharpens, promoting deterministic selection of high-probability actions; conversely, $\tau > 1$ smooths the distribution, yielding more uniform action probabilities. This adjustment enhances sampling diversity, which is critical in adversarial contexts to avoid detection by traffic monitoring systems that flag repetitive or overly deterministic perturbations. By tuning $\tau$, the policy balances exploration and exploitation, optimizing both stealth and adaptability.


\subsection{\textbf{Optimization}}

We employ an actor-critic optimization strategy. The actor, parameterized by $\theta$, maps states to a probability distribution 
$\pi_{\theta}(s_t)$ over actions. The objective of the actor is to select actions that maximize the cumulative future reward:

\begin{equation}\label{equation-16}
\max_{\theta}{\mathbb{E}_{\tau \sim p_{\theta}(\tau)}}[\sum\limits_ {t=1}^T{r(s_t, a_t)}].
\end{equation}

The solution to the above problem can be obtained through iterative updates:

\begin{equation}\label{equation-16}
\theta_{k+1} = \theta_{k} + {\gamma}\mathbb{E}_t[\nabla_{\theta}{\log\pi_{\theta}(a_t|s_t)}{Q^{\pi}(s_t, a_t)}],
\end{equation}
where $\gamma$ represents the step size, and $Q^{\pi}(s_t, a_t)$ is the state-action value function, which indicates the value of taking action $a_t$ in state $s_t$, used to evaluate the quality of the action. However, in practice, two challenges arise: first, the approximate value of $Q$ suffers from high variance; second, the original sampled data cannot be reused, thus requiring resampling for each iteration, which slows down the model update process. To address these issues while preserving the unbiased nature of the target, a baseline is subtracted from $Q$, and the importance sampling method is employed to transform the online policy gradient method into an offline policy gradient method \cite{DBLP:journals/corr/SchulmanWDRK17}:

\begin{equation}
     \mathcal{L}_{policy} = \mathbb{E}_t[\nabla_{\theta}{\frac {\pi_{\theta}(a_t|s_t)} {\pi_{\theta^{\prime}}(a_t|s_t)}}{\log\pi_{\theta}(a_t|s_t)}{A^{\theta^{\prime}}(s_t, a_t)}]
\end{equation}

\begin{equation}\label{equation-16}
\theta_{k+1} = \theta_{k} + {\gamma}L_{policy},
\end{equation}
where $\frac {\pi_{\theta}(a_t|s_t)} {\pi_{\theta^{\prime}}(a_t|s_t)}$ represents the update amplitude of the policy, and $A^{\pi}(s_t, a_t)$ is the advantage function, which is used to measure the relative advantage of the current state-action pair compared to the baseline action. The advantage function is calculated as follows:

\begin{equation}
\begin{aligned}
    A^{\theta^{\prime}}(s_t, a_t) = Q_{\pi_{\theta^{\prime}}}(s_t, a_t)-V_{\pi_{\theta^{\prime}}}(s_t),
\end{aligned}
\end{equation}
where $V_{\pi_{\theta^{\prime}}}(s_t)$ represents the average value of the state $s_t$ under the policy $\pi_{\theta^{\prime}}$.

To further enhance the stochasticity of the policy and improve its exploratory capability, we incorporate entropy regularization into the computation of the policy objective function. By introducing an entropy term, the model is encouraged to maintain a higher degree of randomness during training, thereby mitigating the risk of premature convergence to suboptimal local minima.

\begin{equation}
     \mathcal{L}_{entropy} = \mathbb{E}_t[-\nabla_{\theta}{\pi_{\theta}(a_t|s_t)\log\pi_{\theta}(a_t|s_t)}]
\end{equation}

\begin{equation}\label{equation-16}
\theta_{k+1} = \theta_{k} + {\gamma}\mathcal{L}_{policy} - {\alpha}\mathcal{L}_{entropy},
\end{equation}

The critic network, parameterized by $\phi$, estimates the state value. It approximates the state value by minimizing the Mean Squared Error between the estimated values and the discounted future rewards.
\begin{equation}\label{equation-21}
\min_{\phi}{\mathbb{E}_{t}[(V_{\phi}(s_t)-E_{a \sim \pi_{\theta^{\prime}}}(Q(s_t,a_t)))^2]}.
\end{equation}



\section{Experiment Evaluation}\label{sec:Experiment}



In this section,  multi-facet experiments are conducted to answer the following research questions: 

\textbf{RQ1}: Can the \our effectively generate adversarial traffic under both white-box and black-box constraints? (Section \ref{performance}) 

\textbf{RQ2}: To what extent does the generated adversarial traffic demonstrate transferability across diverse classifiers?  (Section \ref{sec:Transferability})

\textbf{RQ3}: How much does each module of \our contribute to the
model performance? (Section \ref{ablationStudy})



\textbf{RQ4}: How sensitive is \our to hyper-parameters? (Section \ref{sensitivityStudy})  

\textbf{RQ5}: Is \our easy to deploy? (Section \ref{deployment})







\begin{table}[hbt]
	\centering
	\caption{Dataset statistics}
	\label{dataset}
	\setlength{\tabcolsep}{3.5mm}
\begin{tabular}{|c|c|c|c|}
\hline
DataSet    & \#Packet & \#Flow & \#Label \\ \hline
CSTNET-TLS & 178681   & 847  & 118     \\ \hline
ISCX-VPN   & 60000    & 447   & 12      \\ \hline
ISCX-Tor   & 60000    & 252   & 12      \\ \hline
\end{tabular}
\end{table}

\subsection{\textbf{Experimental Settings}} 

\noindent\textbf{Datasets.} To evaluate the effectiveness of our proposed method \our, we conduct extensive experiments across three publicly available datasets. Dataset statistics are summarized in Table~\ref{dataset}. The first dataset, CSTNET-TLS \cite{lin2022bert}, comprises 118 applications collected under CSTNET from March to July 2021. These applications were drawn from the Alexa Top-5000 and utilize TLS 1.3. The second dataset, ISCX-VPN \cite{gil2016characterization}, was captured by the Canadian Institute for Cybersecurity and includes traffic from both VPN and non-VPN environments. It spans six service categories: Chat, VoIP, P2P, Email, File Transfer, and Streaming. The third dataset, ISCX-Tor, released by the University of New Brunswick (UNB), is of particular relevance for analyzing traffic within anonymous communication systems. This dataset obscures user behavior by routing communication through The Onion Router (Tor) and, like ISCX-VPN, includes six service types in both Tor and non-Tor traffic.




\noindent\textbf{Data Pre-processing.} Similar to ET-BERT\cite{lin2022bert}, this study performs a series of preprocessing on the original dataset prior to the experiment. First, the dataset is cleaned by removing the unrelated traffic to the specific protocol transmission, such as the datagrams associated with Address Resolution Protocol (ARP) and Dynamic Host Configuration Protocol (DHCP). To mitigate the interference of identifying information in the datagram header, such as IP addresses and port numbers, the Ethernet header, IP header, and TCP header’s port numbers are discarded. Additionally, we exclude data packets that are either too short or lack a payload. Short data packets, typically used only to establish a connection between the client and the server, do not provide valuable information for classification. Each dataset is divided into training set, validation set and test set in a ratio of 8:1:1.





\noindent\textbf{Traffic Classifiers.} We adopt ET-BERT \cite{lin2022bert}, a state-of-the-art pre-trained traffic classification model, as the baseline for evaluating \our. 


In addition, to further assess the transferability of traffic perturbations, we employ a suite of state-of-the-art traffic classification models. We follow the original implementations to implement these attacks.

\textbf{SVM} \cite{DBLP:journals/tifs/TaylorSCM18} represents a traditional machine learning approach to traffic classification, where the model is trained on manually engineered statistical features extracted from raw traffic. By mapping these handcrafted features into a higher-dimensional space, SVMs attempt to construct decision boundaries that separate different traffic classes.


\textbf{CNN} \cite{7899588} leverages convolutional and pooling operations to automatically extract local discriminative features from traffic data. By sliding convolutional filters over packet byte sequences, CNNs can capture spatially localized patterns. To address the issue of variable-length packet sequences, inputs are typically truncated or padded to a fixed length prior to training.



\textbf{LSTM}\cite{DBLP:conf/ndss/RimmerPJGJ18} is a multi-layer recurrent neural network capable of processing network packets of arbitrary length. Designed to capture long-range dependencies, LSTM is well-suited for modeling the sequential relationships within packet byte streams.

\textbf{Yatc} \cite{DBLP:conf/aaai/0001ZDWWGX23} is a Transformer-based traffic classification framework that integrates Masked Autoencoder (MAE) pre-training with a Multi-level Feature Refinement (MFR) module. During the pre-training phase, YaTC leverages large volumes of unlabeled traffic to learn generic latent representations via the MAE paradigm, thereby capturing structural regularities and contextual dependencies within traffic. Subsequently, in the fine-tuning stage, a small set of labeled traffic samples is used to adapt these representations to specific classification tasks.


\textbf{NetMamba} \cite{DBLP:conf/icnp/WangXWWZ024} introduces a pre-trained state space model tailored for network traffic classification, designed to balance efficiency and accuracy. Built upon the unidirectional Mamba architecture, NetMamba offers a lightweight alternative to Transformer-based models by leveraging state space representations for sequence modeling.




It is important to note that our approach perturbs the content of network packets. Therefore, we focus exclusively on classifiers that utilize packet content as input. Classifiers that rely on metadata features such as packet direction or size, e.g., Exosphere\cite{DBLP:conf/ccs/FuL0024}, DF\cite{sirinam2018deep}, CUMUL\cite{DBLP:conf/ndss/PanchenkoLPEZHW16}, FlowPrint\cite{DBLP:conf/ndss/EdeBCRDLCSP20}, and FS-Net\cite{liu2019fs}, are excluded from consideration.






\noindent\textbf{Adversarial Perturbation Benchmarks.} Since countering the transformer-based pre-trained classifiers is a new task, three  benchmarks are utilized to execute adversarial perturbations on packets:


\begin{itemize}

\item
\textbf{Ditto\cite{meier2022ditto}:} Ditto is a system that provides WAN traffic obfuscation at line rate by transforming real network traffic into a predefined, fixed pattern of packet sizes and timings, making it difficult for eavesdroppers to perform traffic analysis. It achieves this by operating on programmable network switches, using techniques like packet padding, buffering, and chaff packet insertion to hide the underlying traffic's characteristics while running at high speeds with minimal overhead. In this experiment, we modify the data packet to the same length, which is 1500 bytes. 

\item
\textbf{Blind Adversarial Perturbation (BAP)\cite{nasr2021defeating}:} BAP aims to train a generator-like neural network to bypass DNN-based traffic analysis classifiers. It operates by perturbing the characteristics of real-time network flows. The key advantage is its ability to function without the need to cache the flow or pre-know the network flow characteristics, thereby offering greater adaptability. It is important to note that BAP represents a broader class of techniques that modify traffic characteristics by inserting byte sequences at the end of network packets\cite{holland2023detorrent,shenoi2023ipet}.



\item
\textbf{Random Post Padding Perturbation (RandPostPad):} This baseline perturbs packets by appending a randomly generated byte sequence to the packet payload. The appended sequence has fixed length and does not depend on packet semantics or model feedback.

\item
\textbf{RL Post Padding Perturbation (RLPostPad):} To verify the effectiveness of the pre-padding method proposed in this study, we modify the semantic representation of traffic by appending a byte sequence to the end of the packet. This inserted byte sequence is generated using reinforcement learning.

\item
\textbf{Pre-random Padding Perturbation (\our(PRPP)):} This method represents a variant of our proposed approach, which randomly generates byte sequences to modify packet header fields or insert them at the beginning of network packets.

\end{itemize}

\noindent\textbf{Evaluation Scenarios.} This study first evaluates two distinct scenarios: white-box and black-box defenses, as outlined in Section~\ref{DefenseModel}. Subsequently, adversarial transferability is assessed, followed by ablation experiments and parameter sensitivity analysis. Unless otherwise specified, all subsequent experiments adhere to the default untargeted adversarial defense setting introduced in Section~\ref{DefenseModel}, with the padding length fixed at 32 bytes.







\noindent\textbf{Evaluation Metrics.}\label{sec:evalution}


To assess the effectiveness of our approach, we adopt classification accuracy (ACC) as the primary evaluation metric. A lower ACC indicates a stronger defense, as it reflects a higher proportion of adversarial samples that successfully mislead the classification model. Formally, for a given perturbation generator $G$ and test dataset $D_{test}$, ACC is defined as:


\begin{equation}
\begin{aligned}
   \text{ACC} = 1-\frac{|\{x \in D_{\text{test}} : f(G(x)) \neq f(x)\}|}{|D_{\text{test}}|},
\end{aligned}
\end{equation}

where \( f(\cdot) \) denotes the classification model.

\noindent\textbf{Implementation.}
All the experiments are implemented in Python 3.9 and PyTorch 2.3 on a workstation with GPU NVIDIA H100 and Ubuntu 18.04 operating system. In our model-training course, the AdamW optimizer is used to benefit the learning execution. The learning rates for the actor and critic are set to $1\times{10}^{-5}$ and $1\times{10}^{-4}$, respectively. The batch size is set to 32. During the training phase, we conducted one epoch per dataset. Empirical verification across multiple experiments reveals that the model parameters saved after the final training iteration consistently yield the strongest adversarial effect.



\subsection{\textbf{Performance Evaluation}}\label{performance}

To answer \textbf{RQ1}, we evaluate the degradation in traffic classification accuracy on adversarial traffic relative to clean traffic. The evaluation is conducted under both white-box and black-box settings.



\subsubsection{\textbf{Packet Perturbation under Black-Box Defense}}\label{pakcetsBlackBox}

\begin{table}[]
	\centering
	\caption{Classification Accuracy of Network Packets Under Black-Box Settings with Different Adversarial Perturbation}
	\label{accuracyforPackets}
\begin{tabular}{|l|c|c|c|}
\hline
\textbf{Attack Model}              & \textbf{ISCX-VPN} & \textbf{ISCX-Tor} & \textbf{CSTNET-TLS} \\ \hline
No-Defense                         & 0.9943            & 0.9993            & 0.9984              \\ \hline
RandPostPad                        & 0.9935            & 0.9992            & 0.9982              \\ \hline
Ditto                              & 0.9193            & 0.9990            & 0.9984              \\ \hline
RLPostPad                          & 0.9187            & 0.9992            & 0.9983               \\ \hline
BAP                                & 0.8993            & 0.9912            & 0.9843              \\ \hline
\textbf{\ourNoSpace(PRPP)} & \textbf{0.3328}   & \textbf{0.1512}   & \textbf{0.2259}     \\ \hline
\textbf{\our}       & \textbf{0.2568}   & \textbf{0.1367}   & \textbf{0.2025}     \\ \hline
\end{tabular}
\end{table}

We first evaluate the performance of our defense in a black-box scenario. To establish a baseline, we fine-tuned the model on all datasets using the same parameters as ET-BERT \cite{lin2022bert}. As shown in Table~\ref{accuracyforPackets}, without any defense (unperturbed packets), the classification model achieves near-perfect accuracy on all three datasets, closely matching the performance of the original ET-BERT. Based on Table~\ref{accuracyforPackets}, we draw the following observations: i) Compared with RandPostPad, Ditto, RLPostPad, and BAP, our method \ourNoSpace(PRPP) achieves substantial improvements, reducing classification accuracy by 56.65\%, 84.0\%, and 75.84\% on the ISCX-VPN, ISCX-Tor, and CSTNET-TLS datasets, respectively, thereby significantly outperforming existing defense techniques. This improvement is attributed to our pre-padding perturbation strategy, which alters the semantic representation of network traffic and disrupts the classifier’s learned features. By contrast, Ditto, RLPostPad, and BAP employ post-padding strategies that modify packets without affecting semantics;  ii) \our further benefits from reinforcement learning-based optimization. Relative to basic PRPP, the classification accuracy on the three datasets is further reduced by 7.6\%, 1.45\%, and 2.34\%, respectively. This improvement stems from the model’s ability to generate traffic-specific perturbation sequences, resulting in stronger adversarial effects; iii) The results of RandPostPad, RLPostPad and BAP show that classification accuracy across the three datasets remains above 89\%, indicating that even with generation mechanisms, post-padding strategies fail to produce sufficiently effective perturbations against pre-trained classifiers. This further underscores the effectiveness of our proposed approach.

\subsubsection{\textbf{Burst Perturbation under Black-box Defense}}\label{burstBlack}

\begin{table}[]
	\centering
	\caption{Classification accuracy on network flow using different perturbation manners Under Black-Box Settings}
	\label{accuracyforBurstUnderBlack}
\begin{tabular}{|l|c|c|c|}
\hline
\textbf{Attack Model}              & \textbf{ISCX-VPN} & \textbf{ISCX-Tor} & \textbf{CSTNET-TLS} \\ \hline
No-Defense                         & 0.9217            & 0.7500            & 0.6045              \\ \hline
RandPostPad                        & 0.7942            & 0.1984            & 0.6068              \\ \hline
Ditto                              & 0.7673            & 0.2024            & 0.5891              \\ \hline
RLPostPad                          & 0.8345            & 0.3452            & 0.5939              \\ \hline
BAP                                & 0.8256            & 0.3654            & 0.5721               \\ \hline
\textbf{\ourNoSpace(PRPP)} & \textbf{0.6286}   & \textbf{0.1230}   & \textbf{0.1074}     \\ \hline
\textbf{\our}       & \textbf{0.4899}   & \textbf{0.1090}   & \textbf{0.0390}     \\ \hline
\end{tabular}
\end{table}

Network traffic can be represented not only at the packet level but also as a transmission-guided structure (BURST), which serves as an alternative input to traffic classifiers. To further evaluate the effectiveness of our method, we apply adversarial perturbations to BURST by deploying a perturbation model trained on packet structures to each packet within BURST, thereby modifying its semantics. The experimental results are presented in Table~\ref{accuracyforBurstUnderBlack}, from which we draw the following observations: i) Similar to packet-level perturbations, \our consistently outperforms all baselines across all datasets. Compared with RandPostPad, Ditto, RLPostPad, and BAP, it reduces classification accuracy by 27.74\%, 8.94\%, and 53.31\% on the ISCX-VPN, ISCX-Tor, and CSTNET-TLS datasets, respectively; ii) A comparison between \ourNoSpace(PRPP) and \our highlights the critical role of reinforcement learning, which further reduces classification accuracy by up to 13.87\% across the three datasets; iii) Despite introducing perturbations into every packet, RandPostPad, BAP, and RLPostPad still exhibit poor performance, particularly on the ISCX-VPN and CSTNET-TLS datasets. This suggests that post-padding operations are insufficient to alter the BURST-level semantic representations captured by advanced classifiers, whereas our method effectively disrupts these representations and achieves substantial defense gains.

\subsubsection{\textbf{Packet Perturbations under White-Box Defense}}\label{pakcetsWhite}

\begin{figure}[hbt]
	\centering
        \includegraphics[scale=0.35]{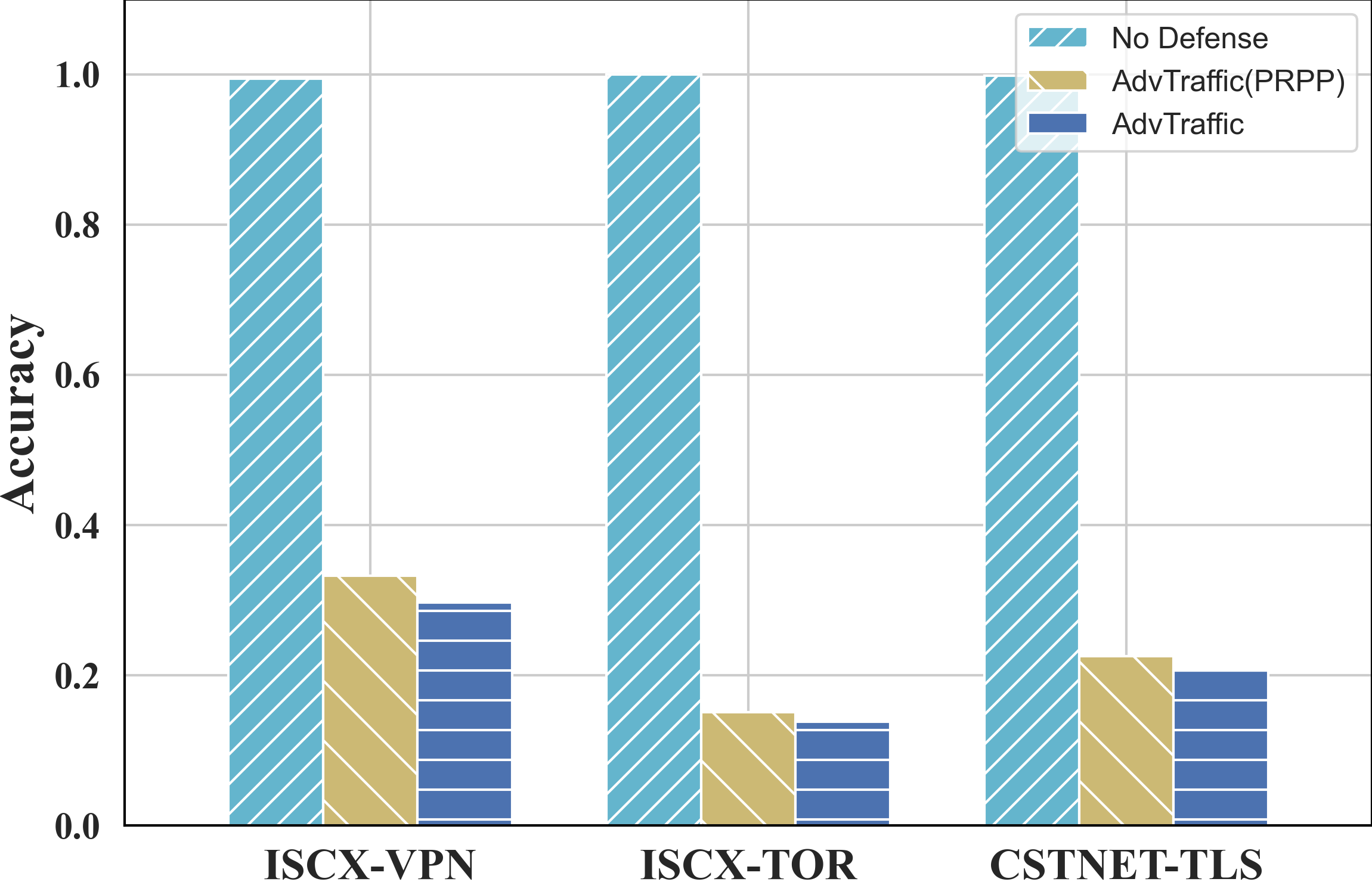}
	\caption{Performance with packet perturbations under White-box defense.}
	\label{fig:whiteboxpacket}
\end{figure}

As previously discussed, the baseline perturbation methods underperform compared to \our in the black-box scenario. Therefore, to conserve space, we focus our evaluation on the effectiveness of \our in the white-box setting. The results with a padding length of 32 bytes, shown in Figure~\ref{fig:whiteboxpacket}, demonstrate that \our also remains effective. Two key observations can be drawn. First, across all three datasets, packet perturbation substantially reduces classification accuracy: by 70.88\%, 90.08\%, and 80.65\% on ISCX-VPN, ISCX-Tor, and CSTNET-TLS, respectively. Second, reinforcement learning optimization further enhances defense effectiveness by generating packet-specific perturbations. Compared with \ourNoSpace(PRPP), classification accuracy is additionally reduced by 4.73\%, 5.27\%, and 3.40\% on ISCX-VPN, ISCX-Tor, and CSTNET-TLS, respectively.

\subsubsection{\textbf{Burst Perturbation under White-box Defense}}\label{burstWhite}

\begin{figure}[hbt]
	\centering
        \includegraphics[scale=0.35]{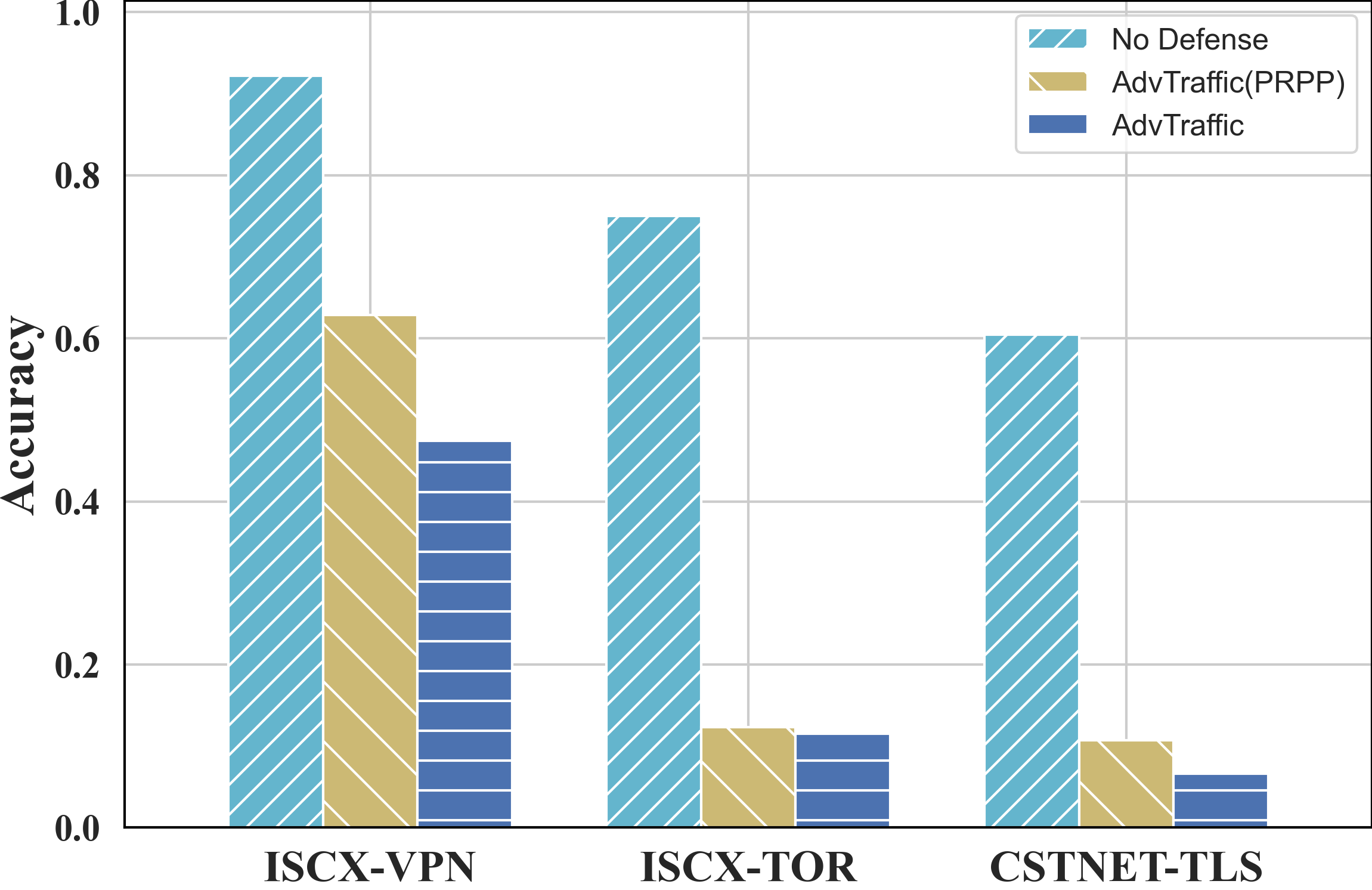}
	\caption{Classification accuracy on network flow using different perturbation manners Under White-box Settings.}
	\label{fig:whiteboxflow}
\end{figure}

The experimental results for BURST in the white-box scenario are presented in Figure~\ref{fig:whiteboxflow}. Two key observations can be made. First, across all three datasets, packet perturbations remain highly effective in reducing classifier accuracy; for instance, on the CSTNET-TLS dataset, accuracy drops to as low as 6.61\%. Second, reinforcement learning further enhances perturbation generation, yielding additional reductions in accuracy—by up to 15.43\% compared with \ourNoSpace(PRPP).

In summary, our proposed network traffic perturbation method effectively defends against both packet-level and burst-level traffic classification. By leveraging reinforcement learning, it ensures robust perturbation generation in both black-box and white-box settings. These results demonstrate the strong effectiveness of our approach in adversarially perturbing network traffic.

\subsection{\textbf{Adversarial Perturbation Transferability}}\label{sec:Transferability}


To address \textbf{RQ2}, we investigate the transferability of adversarial perturbations, specifically whether parameters trained on ET-BERT can be applied to other classifiers without retraining while still achieving effective adversarial effects. For each classifier, we evaluate both clean samples and adversarially perturbed samples. 

The results in Table~\ref{tab:transftable} yield several key insights: i) YATC, a Transformer-based traffic classifier, experiences an average accuracy reduction of 21.5\% on the ISCX-VPN, ISCX-Tor and CSTNET-TLS datasets, confirming the transferability of adversarial traffic. This is attributable to the shared Transformer architecture in both ET-BERT and YATC, which similarly capture contextual dependencies in traffic. In contrast, NetMamba, which adopts the Mamba architecture instead of the Transformer, still suffers an average accuracy drop of 34.72\%, indicating that adversarial perturbations can generalize across different backbone architectures; ii) Despite substantial architectural differences, 1D-CNN and LSTM classifiers also exhibit significant performance degradation, with average accuracy reductions of 17.75\% and 39.1\%, respectively. This demonstrates that adversarial traffic transfers effectively across diverse deep learning models, disrupting classifier performance; iii) Interestingly, even traditional SVM classifiers, which rely on traffic-level statistical features, are affected. Although only a 32-byte perturbation is introduced—representing a negligible fraction of the original traffic and intuitively unlikely to alter statistical features—the average accuracy still decreased by 8.71\%. This result highlights the broad adversarial effectiveness of our method, extending beyond deep learning-based models.

\begin{table*}[t]
\centering
\caption{Accuracy of Various Attacks on Defense Datasets}
\label{tab:transftable}
\footnotesize
\setlength{\tabcolsep}{6pt}
\renewcommand{\arraystretch}{1.2}
\begin{tabular}{l|c|c|c|c|c|c} 
\toprule
\multirow{2}{*}{\textbf{Attack Model}} & 
\multicolumn{2}{c|}{\textbf{ISCX-VPN}} & 
\multicolumn{2}{c|}{\textbf{ISCX-Tor}} & 
\multicolumn{2}{c}{\textbf{CSTNET-TLS}} \\
\cmidrule(lr){2-3} \cmidrule(lr){4-5} \cmidrule(lr){6-7}
& \textbf{No Defense} & \textbf{\our} 
& \textbf{No Defense} & \textbf{\our} 
& \textbf{No Defense} & \textbf{\our} \\
\midrule
SVM        & 0.9066 & \textbf{0.7986} & 0.3836 & \textbf{0.2818} & 0.3003 & \textbf{0.2488} \\
\hline 
1D-CNN     & 0.9928 & \textbf{0.8315} & 0.9957 & \textbf{0.7682} & 0.7750 & \textbf{0.6313} \\
LSTM       & 0.9948 & \textbf{0.6053} & 0.9910 & \textbf{0.7703} & 0.8815 & \textbf{0.3187} \\
\hline 
YATC       & 0.5871 & \textbf{0.4598} & 0.8221 & \textbf{0.7549} & 0.8797 & \textbf{0.4292} \\
NetMamba   & 0.5739 & \textbf{0.3262}      & 0.9993 & \textbf{0.3222} & 0.1546 & \textbf{0.0379} \\
\bottomrule
\end{tabular}
\end{table*}

\subsection{\textbf{Ablation Study}}\label{ablationStudy}

\subsubsection{\textbf{The Impact of Padding Position}}

\begin{figure}[ht]
    \begin{minipage}[b]{0.48\textwidth}
        \centering
        \includegraphics[scale=0.35]{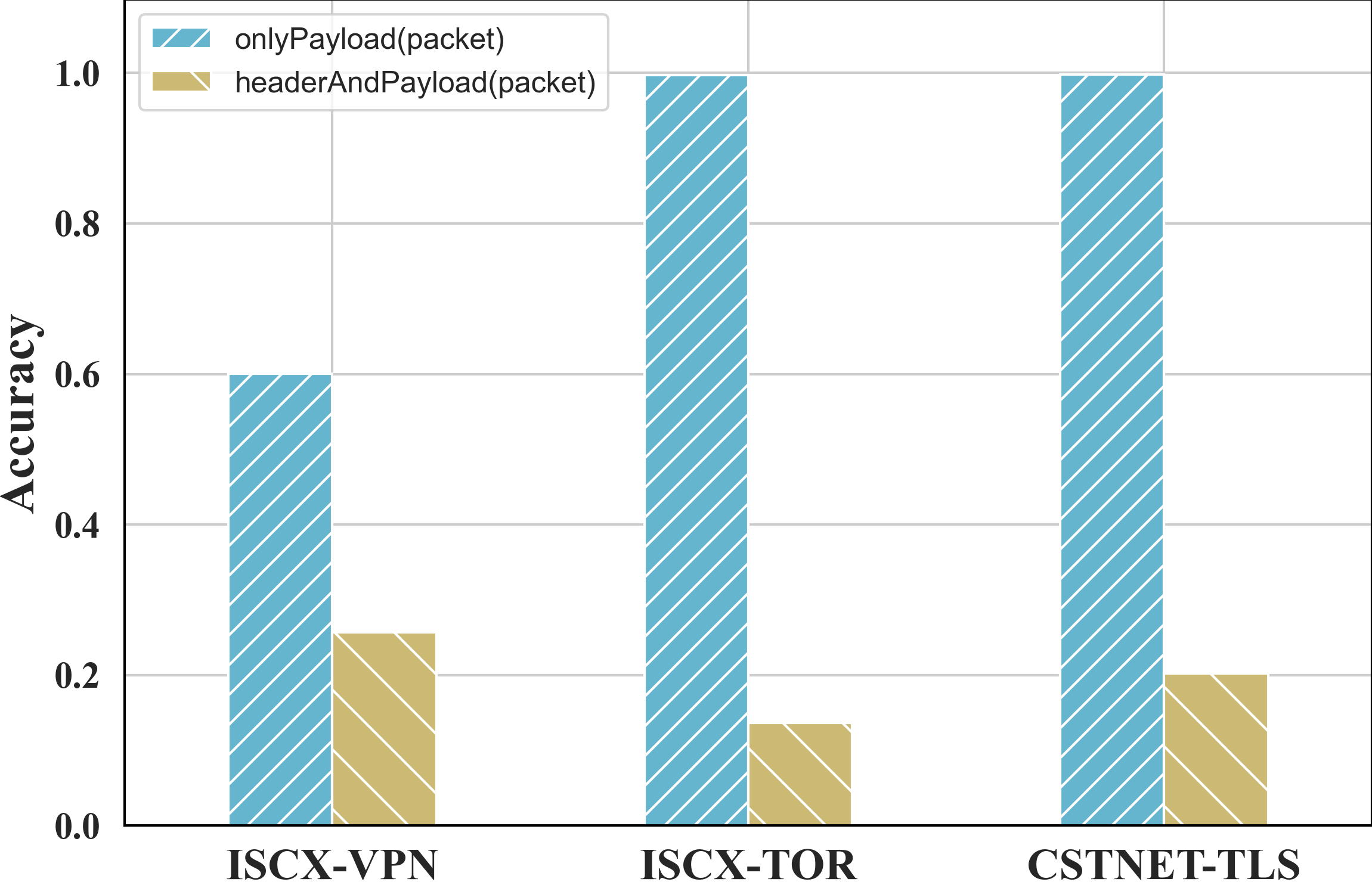}
	\caption{Performance Comparison with and without Packet Header Modification On Packet.}
	\label{fig:onlyPayloadModifyOnPacket}
    \end{minipage}
    \hfill
    \begin{minipage}[b]{0.48\textwidth}
	\centering
	\includegraphics[scale=0.35]{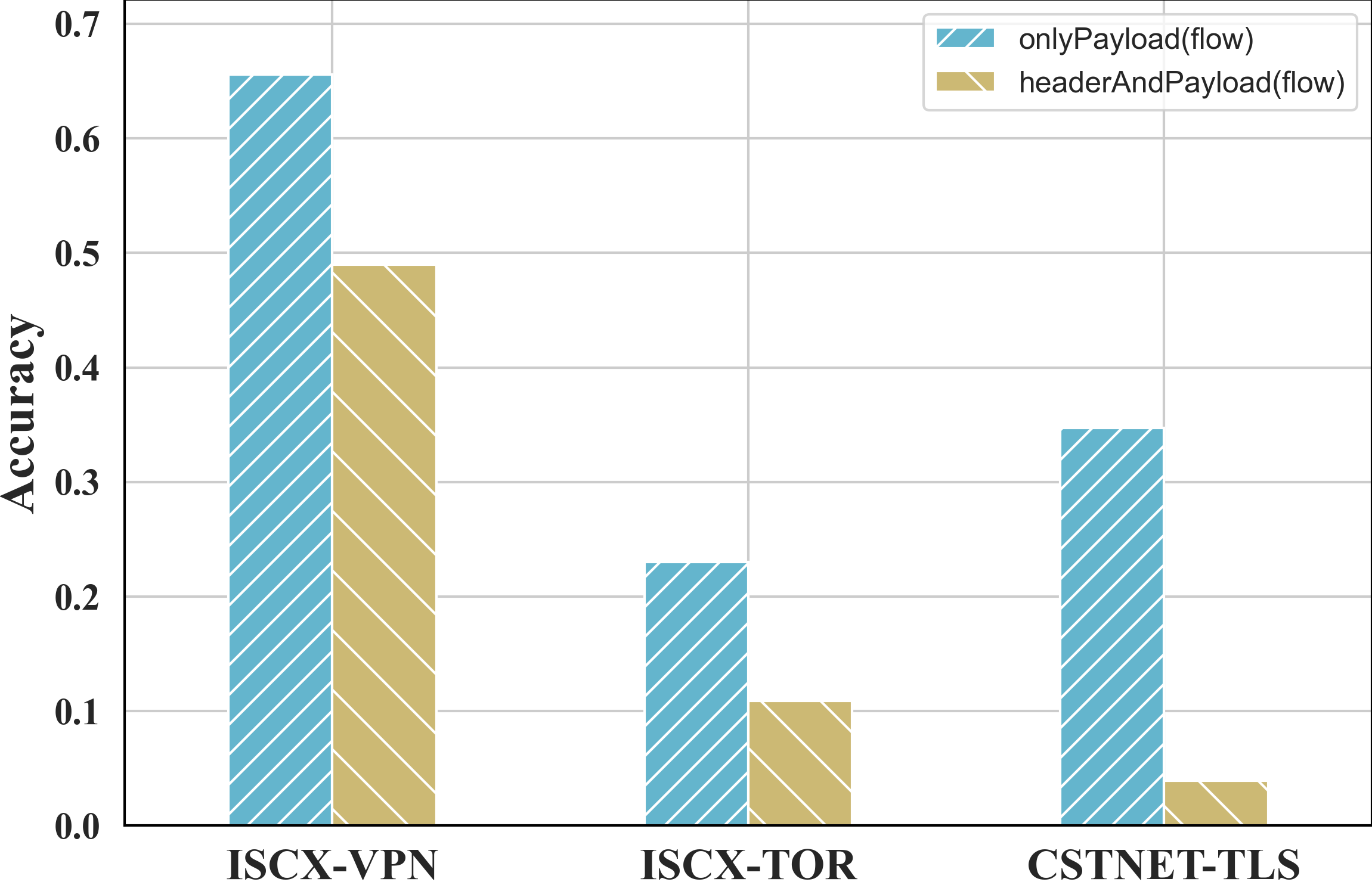}
	\caption{Performance Comparison with and without Packet Header Modification On Burst(Flow).}
	\label{fig:onlyPayloadModifyOnFlow}
    \end{minipage}
\end{figure}

To maximize adversarial effectiveness, we employ a dual strategy that combines modification of specific packet header fields with the insertion of adversarial byte sequences before the packet payload. To isolate the contribution of header modification, we conduct comparative experiments in which only payload-level adversarial bytes are inserted, leaving headers unmodified. These experiments are performed at both packet and burst levels, as illustrated in Figure~\ref{fig:onlyPayloadModifyOnPacket} and Figure~\ref{fig:onlyPayloadModifyOnFlow}. The results show that, across all three datasets, header modification substantially enhances adversarial performance at both levels. Specifically, packet-level classification accuracy decreased by an additional 34.39\%, 86.03\%, and 79.5\% on the ISCX-VPN, ISCX-TOR, and CSTNET-TLS datasets, respectively. At the burst level, accuracy decreased by an additional 16.56\%, 12.12\%, and 30.81\%. These results underscore the necessity of incorporating header-level perturbations to achieve stronger adversarial effects.



\subsubsection{\textbf{The Effect of The Reward Function}}

\begin{figure}[]
	\centering
    \includegraphics[scale=0.35]
    {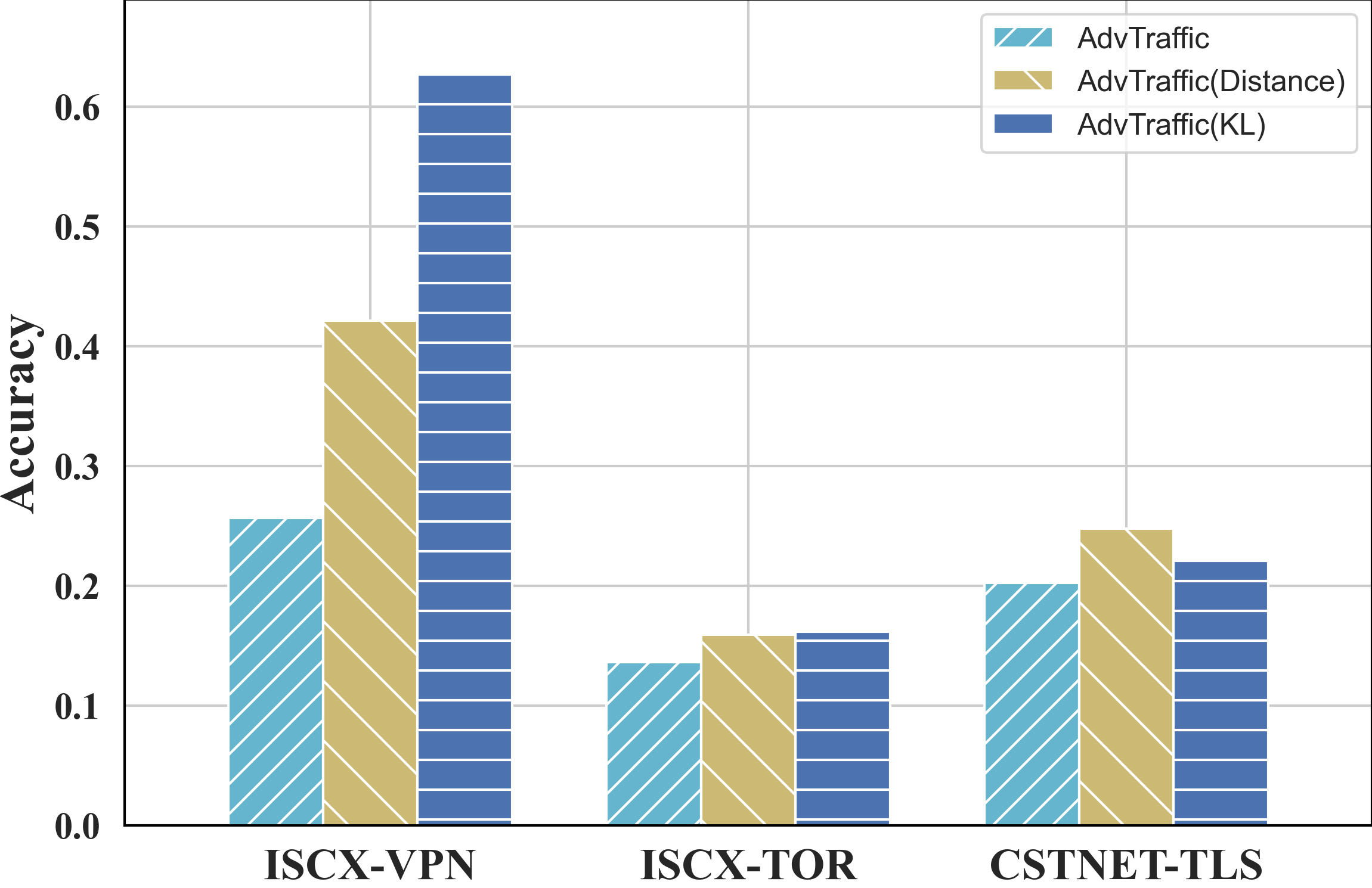}
	\caption{Performance under different reward functions.}
	\label{fig:ablationStudy}
\end{figure}

In the white-box defense scenario, the reward function is designed with two complementary components: one measuring the representation distance, and the other quantifying the dissimilarity between representation distributions. To assess whether this composite design indeed improves learning, we conducted an ablation study in which only one component is retained while the other is discarded. The experimental results, presented in Figure \ref{fig:ablationStudy}, are obtained under a uniform padding length of 32 bytes across all datasets. The findings consistently reveal that using either component in isolation yields inferior performance compared to employing the full reward function. For example, on the ISCX-VPN dataset, classification accuracy drops by 16.45\% when only the distance metric is used and by 36.99\% when only KL divergence is applied. Similar performance degradation was observed on ISCX-Tor and CSTNET-TLS, further confirming that both components are indispensable. These results clearly demonstrate that the integration of both components is essential, and highlight the effectiveness of the proposed composite reward function in enhancing the adversarial perturbation model’s performance across datasets.



\subsection{\textbf{Parameter Sensitivity Study}}\label{sensitivityStudy}

\subsubsection{\textbf{The Effect of Padding Length on Byte Sequences}}\label{sec:paddingLength}

\begin{figure}[]
	\centering
    \includegraphics[scale=0.35]{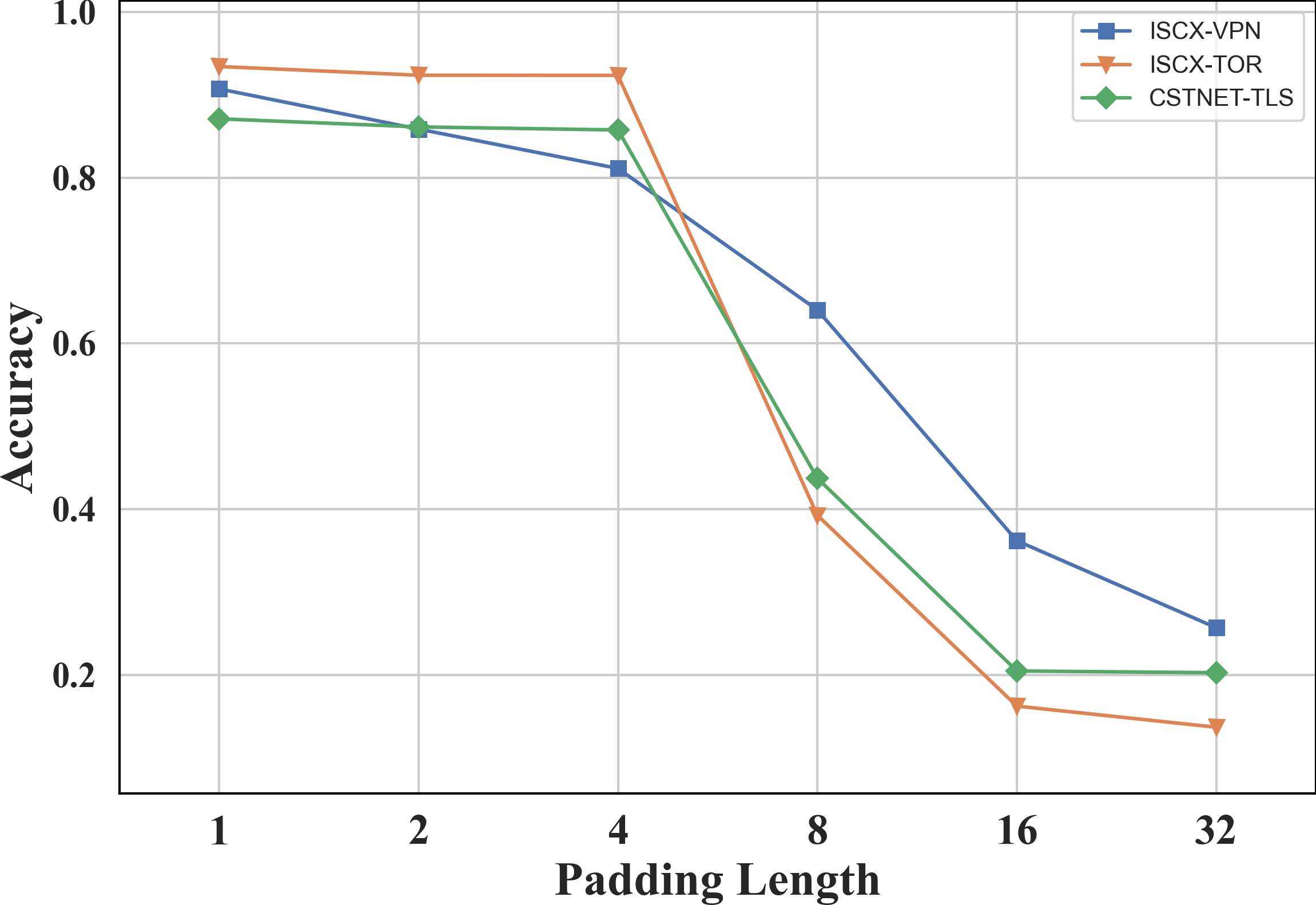}
	\caption{The Effect of Padding Length.}
	\label{fig:paddinglength}
\end{figure}

We further examined how the length of adversarial padding influences perturbation effectiveness. In these experiments, padding lengths of 1, 2, 4, 8, 16, and 32 bytes are applied to packets, with the perturbation model trained under a black-box setting. The experimental results, presented in Figure \ref{fig:paddinglength}, yield several key observations: i) A clear positive correlation emerges between padding length and adversarial strength: longer perturbation sequences consistently induce greater reductions in classification accuracy across all datasets; ii) When the padding length is less than or equal to 8 bytes, classification accuracy still decreases, despite only partial modification of TCP header fields. For example, when the padding length is set to 8 bytes, the classification accuracy decreases by 35.4\%, 60.68\%, and 56.12\% on the ISCX-VPN, ISCX-Tor, and CSTNET-TLS datasets, respectively. This suggests that perturbations confined to the TCP header alone can produce an adversarial effect; iii) A pronounced adversarial effect emerges once the padding length reaches 16 bytes, with classification accuracy reduced to 36.18\%, 16.22\%, and 20.46\% on the ISCX-VPN, ISCX-Tor, and CSTNET-TLS datasets, respectively. This effect arises because both the TCP header and the packet payload are perturbed simultaneously. These findings highlight the necessity of jointly perturbing header and payload fields to substantially enhance defense effectiveness; iv) Beyond 16 bytes, however, the incremental performance gains diminish, reflecting a classic trade-off between adversarial effectiveness and resource overhead. While longer sequences further enrich semantic perturbations, they also incur greater bandwidth consumption. Collectively, these findings underscore the need to select an appropriate padding length that balances adversarial strength against communication efficiency.



\subsubsection{\textbf{The Effect of Temperature Coefficient}}\label{sec:temperature}

\begin{figure}[ht]
    \begin{minipage}[b]{0.48\textwidth}
        \centering
        \includegraphics[scale=0.4]{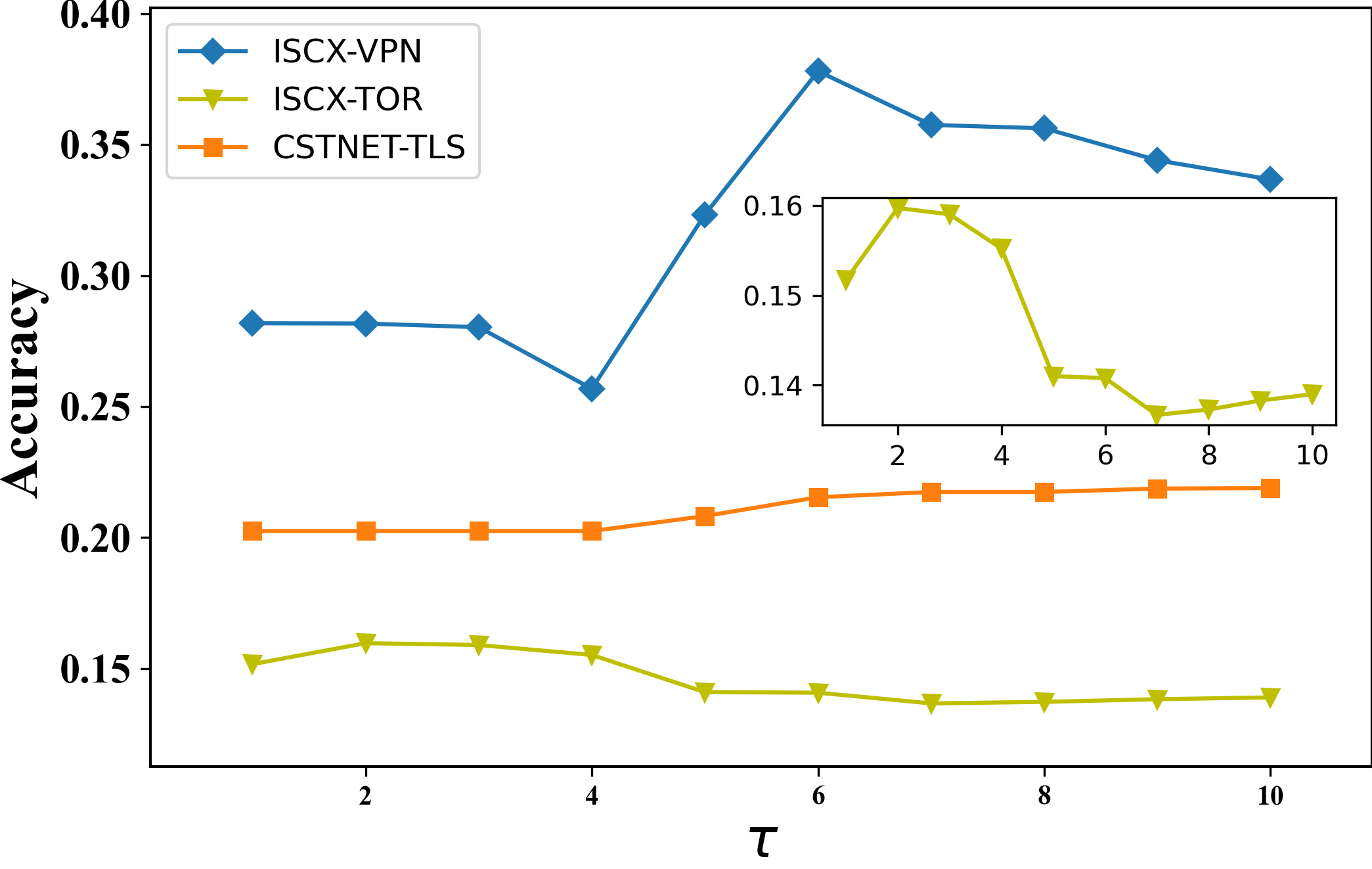}
	\caption{The Effect of Temperature Coefficient.}
	\label{fig:temperature}
    \end{minipage}
    \hfill
    \begin{minipage}[b]{0.48\textwidth}
	\centering
	\includegraphics[scale=0.4]{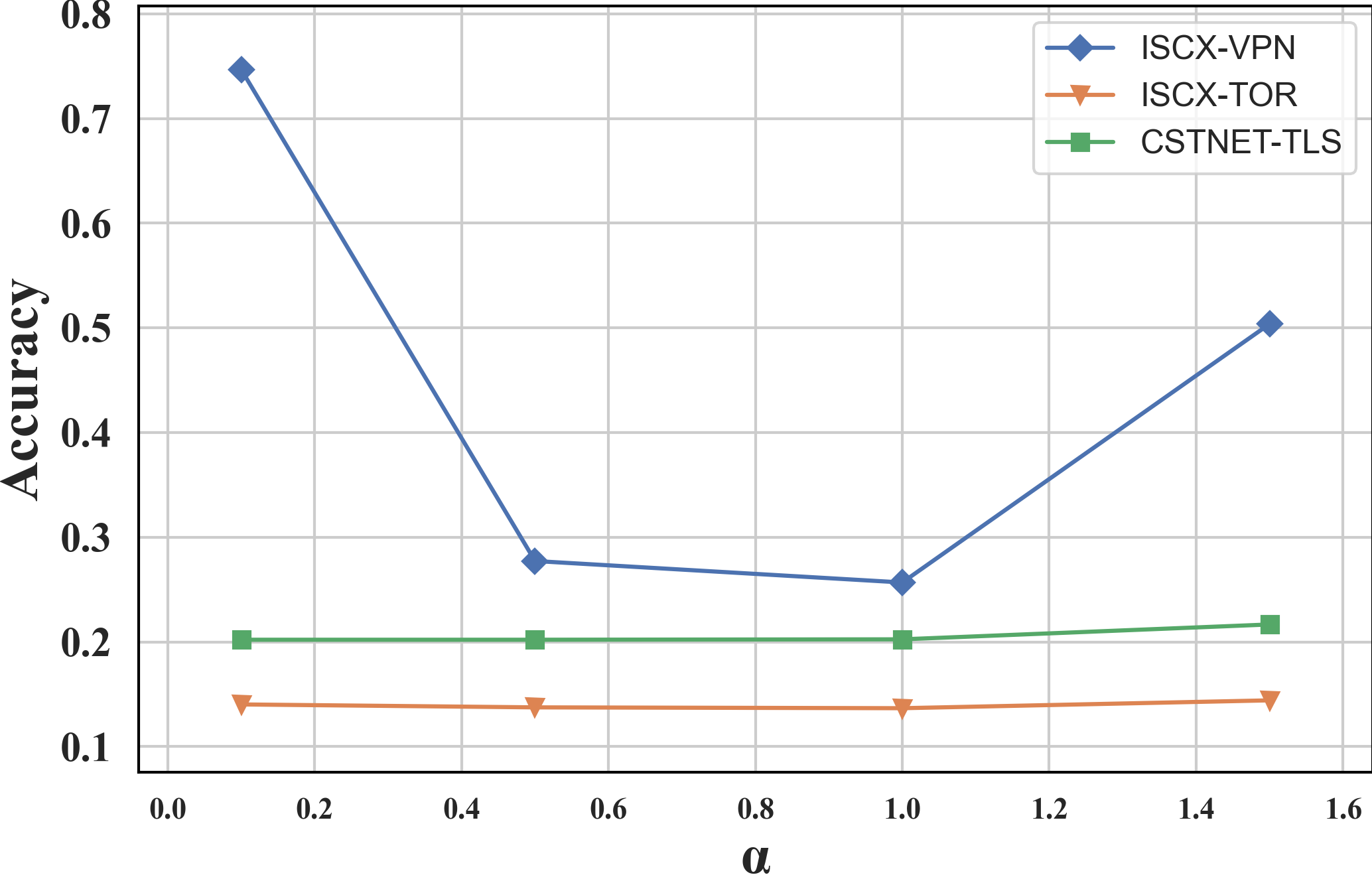}
	\caption{Impact of entropy loss.}
	\label{fig:entropy}
    \end{minipage}
\end{figure}

In adversarial perturbation generation, the temperature coefficient regulates the sharpness of the action probability distribution. We examine its effect on both classification accuracy and action behavior, with the results summarized in Figure \ref{fig:temperature}. Several observations emerge: i) for both the ISCX-VPN and ISCX-Tor datasets, classification accuracy exhibits a non-monotonic trend as the temperature coefficient increases—initially decreasing and then rising. This fluctuation can be attributed to the fact that higher temperature values introduce greater randomness into action selection, thereby destabilizing the convergence of the perturbation strategy. The appendix \ref{appendix:tempetaure} further supports this observation, showing that larger temperature coefficients lead to increasingly uncertain probability distributions; ii) the ISCX-VPN dataset is more sensitive to changes in temperature than ISCX-Tor, with a difference of 12.12\% in adversarial success rate observed between temperature values of 1 and 10; iii) the CSTNET-TLS dataset demonstrates the lowest sensitivity to this parameter, exhibiting only a gradual decline in classification accuracy as the temperature increases. Collectively, these findings underscore the trade-off between adversarial effectiveness and the stochasticity introduced by temperature scaling.


\subsubsection{\textbf{The Impact of Entropy Loss on Model Performance}}

In the policy network’s loss function, the entropy regularization coefficient $\alpha$ plays a pivotal role in shaping model performance. To investigate its influence, we conduct experiments with $\alpha$ values of 0.1, 0.5, 1.0, and 1.5 under a 32-byte padding setting. The results, presented in Figure \ref{fig:entropy}, yield several notable observations. First, the ISCX-VPN dataset is most sensitive to changes in $\alpha$, with the difference between the best and worst classification accuracy reaching 48.99\%. Second, the impact of $\alpha$ on the ISCX-Tor and CSTNET-TLS datasets is far less pronounced, with accuracy differences of only 0.75\% and 1.42\%, respectively. Third, on the ISCX-VPN dataset, increasing $\alpha$ to 1.0 significantly decreases classification accuracy, thereby improving adversarial performance. However, further increases beyond this threshold cause adversarial performance to deteriorate, reflecting an over-regularization effect. Overall, these findings demonstrate that $\alpha$ exerts dataset-specific effects and that careful tuning of this coefficient is essential to achieving optimal adversarial performance.

\subsection{\textbf{Feasibility of Deployment}}\label{deployment}

\subsubsection{\textbf{Caching-based Online Perturbation}}


After validating the effectiveness of \our, we further examine its feasibility for practical deployment. Ideally, the perturbation algorithm should be integrated into the network transport layer to deform packets at line speed. However, the current implementation relies on GPU-based inference, which introduces non-negligible latency. For example, performing four single-step inferences on an NVIDIA H100 GPU incurs an average latency of approximately 100 ms. Although modest in absolute terms, this latency is significant compared to typical packet transmission times at the millisecond or sub-millisecond scale, potentially impacting real-time communication.



To address this challenge, we propose a caching-based strategy. Instead of generating adversarial byte sequences at runtime, pre-generated sequences are derived from the trained perturbation model and stored locally. These sequences are periodically synchronized with a central server to ensure both diversity and freshness. During perturbation, the system randomly selects a sequence from the cache to perturb incoming packets, thereby eliminating the need for performing neural-network inference. This design minimizes computational overhead and guarantees that normal communication remains unaffected.

\begin{figure}[hbt]
	\centering
        \includegraphics[scale=0.35]{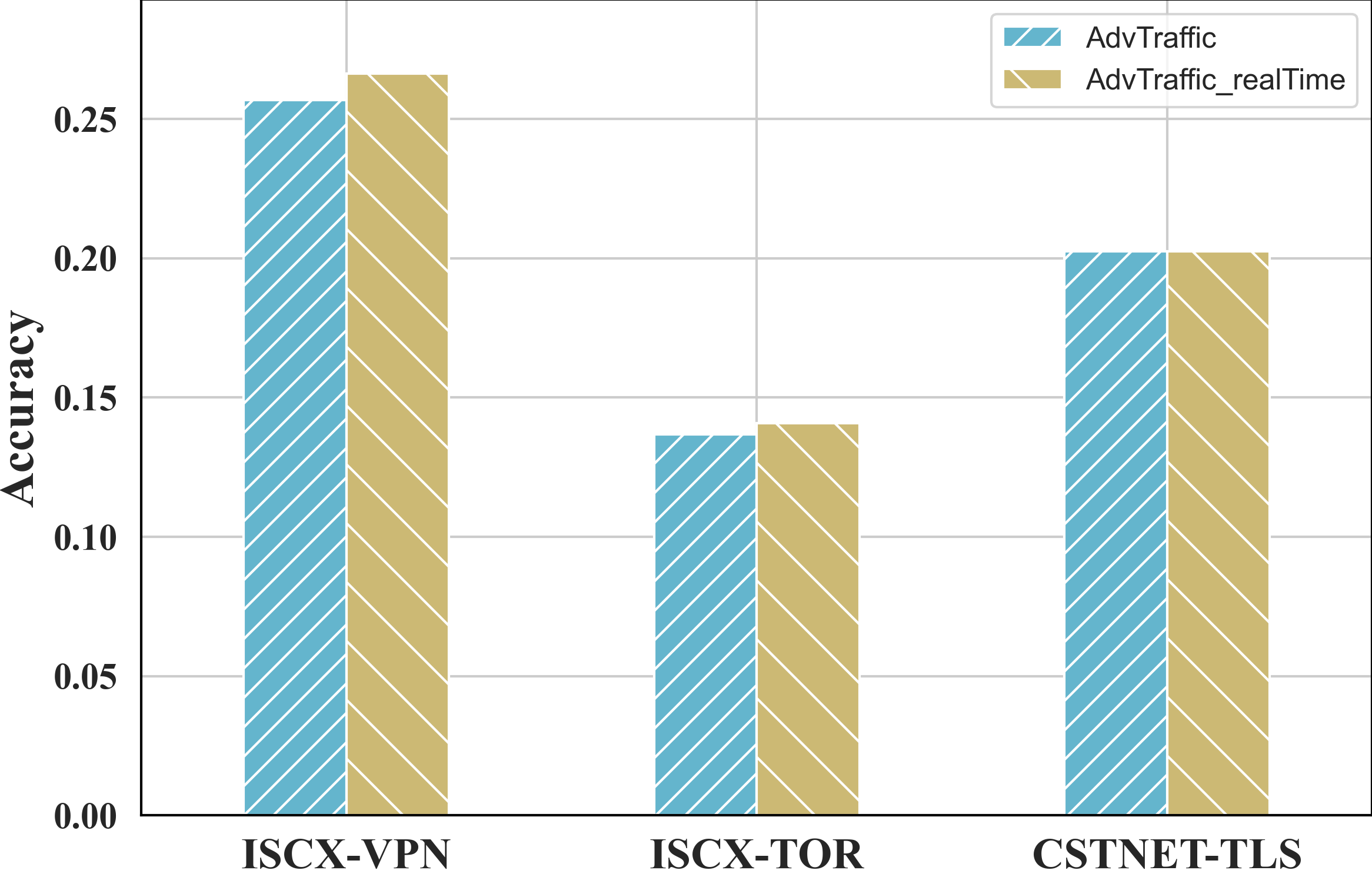}
    
	\caption{Performance with online real-time perturbation.}
	\label{fig:onLineRealTime}
\end{figure}

Figure~\ref{fig:onLineRealTime} illustrates the effectiveness of this strategy. As shown, the classification accuracy degradation caused by online perturbation is comparable to that of per-packet adversarial computation. This consistency demonstrates that the caching-based approach preserves the defense strength while significantly improving efficiency, confirming the practical feasibility of deploying \our in real-world network environments.



\subsubsection{\textbf{Packet Modification Delay}}

In addition, to precisely quantify the time overhead associated with packet modification, we implemented a pluggable module named TrafficPerturb. For seamless integration with the network stack, TrafficPerturb was developed using the eBPF framework\footnote{https://github.com/eunomia-bpf/bpf-developer-tutorial/blob/main/src/20-tc/README.zh.md}, which allows for real-time interception and perturbation of network packets. Experimental results indicate that the average delay introduced per packet is approximately 0.45 milliseconds, suggesting a negligible impact on overall network latency. Similar to other pluggable transport protocols, TrafficPerturb must be deployed at both the client side (i.e., the traffic sender) and at the receiving or egress point (e.g., a gateway).

\subsubsection{\textbf{The Bandwidth Overhead of Traffic Perturbation}}\label{sec:bandwidth}

\begin{table}
\centering
\setlength\tabcolsep{8pt}
\caption{The Ratio of Padding Length to Total Packet Length}
\label{tab:paddingOverhead}
\begin{tabular}{|c|c|c|}
\hline
DataSet    & Mean Length & Ration \\ \hline
ISCX-VPN   & 1106        & 2.9\%  \\ \hline
ISCX-Tor   & 994         & 3.2\%  \\ \hline
CSTNET-TLS & 919         & 3.4\%  \\ \hline
\end{tabular}
\end{table}

Appending byte sequences to network traffic inevitably increases packet size and incurs bandwidth overhead. In bandwidth-constrained environments, excessive padding may lead to congestion or performance degradation. To evaluate this impact, we compute the average packet length across three datasets and quantify the overhead introduced by 32-byte padding. As shown in Table~\ref{tab:paddingOverhead}, the maximum relative increase is only 3.4\%, indicating that the bandwidth overhead introduced by our method is minimal and practically negligible.

\subsubsection{Deployment of \our}\label{appendix:deploy}

\begin{figure}[]
	\centering
	\includegraphics[scale=0.40]{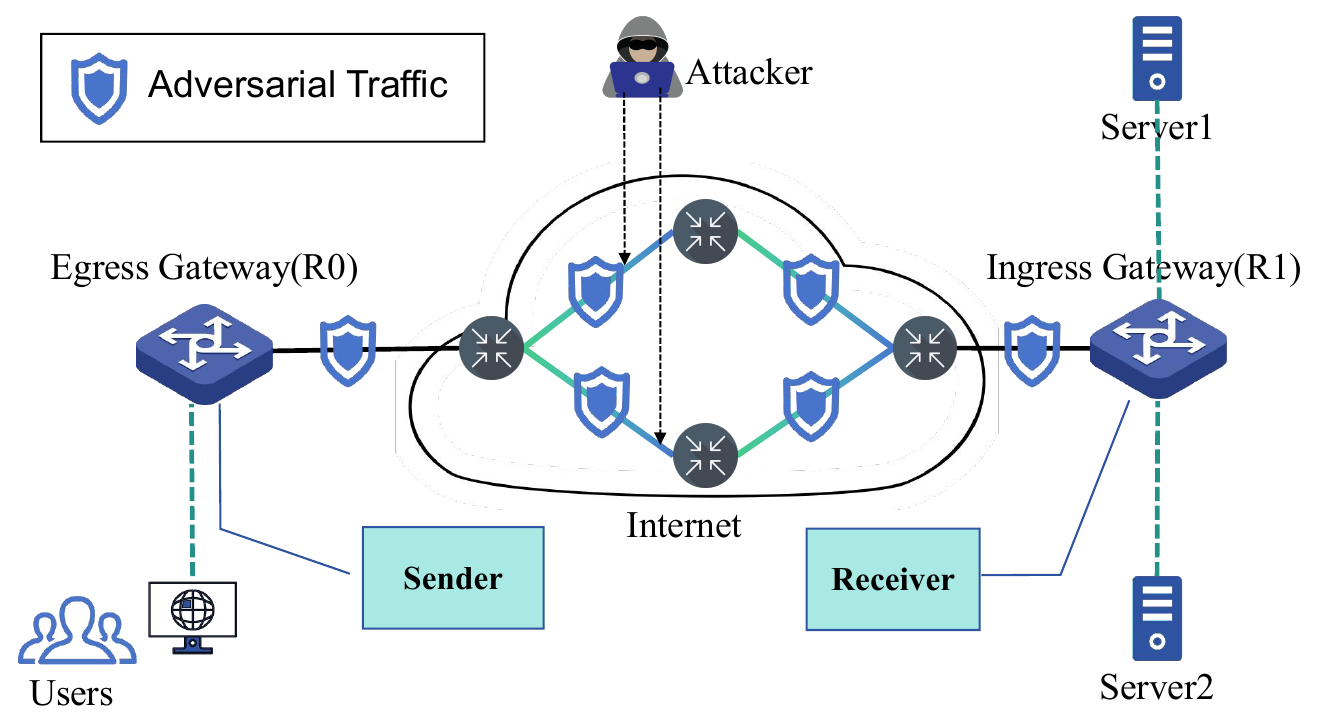}
	\caption{The feasible deployment environment of our \ourNoSpace. The network can be a service provider, a private network that provides content services, etc.}
	\label{fig:deployment}
\end{figure}

The proposed traffic defense method can be applied to networks capable of intercepting and modifying traffic (requiring appropriate permissions). Such networks include carrier networks, overlay networks, and cloud networks. A typical deployment diagram is presented in Figure~\ref{fig:deployment}. The sender (Perturber) module can be deployed at the traffic egress gateway or directly on the host generating traffic, while the receiver module can be placed at the ingress gateway or on the service-providing target machine, such as a server. This deployment strategy ensures that \our remains resilient against manipulation or interference by attackers. It should be noted that actual deployment requires consideration of the controllability of network nodes.


\subsection{\textbf{Discussion}}



Numerous deep learning–based methods have been developed for traffic classification. The adversarial perturbation proposed in this study demonstrates high effectiveness against state-of-the-art transformer-based models. In contrast, graph-based approaches, such as TFE-GNN \cite{zhang2023tfe}, represent fundamentally different paradigms. Investigating the adversarial defense mechanisms of these models remains an important direction for future research.
\section{Related Work} \label{sec:Related}

\subsection{\textbf{Traffic Classification Techniques}}\label{trafficClassTec}




\textbf{Feature-based Methods.}
These methods leverage representative features to model and classify network flows \cite{DBLP:conf/uss/Xue0HE24,9305740, rahman2019tik, shen2019encrypted, taylor2017robust, hayes2016k, al2016adaptive, wang2014effective, cai2012touching, panchenko2011website}. Since an individual packet contains limited information, most feature-based models aggregate statistical features across multiple packets. However, two major challenges exist. First, designing generalizable statistical features to handle increasingly complex and voluminous traffic remains difficult. Second, with the advent of traffic obfuscation techniques \cite{meier2022ditto}, the feature-based methods almost become invalid.


\textbf{Supervised Deep Learning-based Methods.}
This line of studies\cite{Bhat2018VarCNNAD, sirinam2018deep, rimmer2017automated, zha2020meta, zheng2020learning} achieves the automatic extraction on traffic features. However, these DL-based methods require large amounts of labeled data, which is time-consuming and labor-intensive. 



\textbf{Transformer-based Methods.}
The emergence of large language models (LLMs), such as Transformer \cite{devlin2018bert} and BERT variants \cite{lan2019albert, liu2019roberta}, has led to the rise of pre-trained traffic classification methods, with ET-BERT \cite{lin2022bert} as a prominent example. These models significantly outperform traditional approaches and, in theory, can accurately classify traffic from a single IP packet, rendering encryption and obfuscation largely ineffective. To address this challenge, we propose a novel traffic perturbation method to counteract the capabilities of pre-trained classifiers.



\textbf{Graph-based Methods.} Unlike transformer-based NTC methods, this kind of methods first transforms the byte sequence into a graph structure, then employs some techniques, such as contrastive learning, to derive traffic representation. For example, TFE-GNN \cite{zhang2023tfe} constructs byte-level traffic graph to identify potential correlations between raw bytes, and encodes each packet into a representation for traffic classification.



\subsection{\textbf{Traffic Classification Countermeasures}}
Various countermeasure mechanisms \cite{cai2012touching, juarez2015wtf, rahman2020mockingbird, wang2014effective, zhang2019statistical} have been proposed to counter network traffic classification, with traffic obfuscation being one of the most prominent at present. 

\textbf{Mimicry-based Countermeasure.}
These countermeasures disguise traffic by reshaping it to mimic target characteristics \cite{meier2022ditto, mohajeri2012skypemorph}. However, such methods fall short of achieving true unobservability, as shown in \cite{houmansadr2013parrot}.



\textbf{Tunnel-based Countermeasure.}
Tunnel-based methods conceal target traffic by transmitting it through alternative protocols\cite{fifield2015blocking, barradas2018effective, wang2015seeing}. However, these methods are vulnerable to machine learning (ML)-based classifiers \cite{barradas2018effective, wang2015seeing}.


\textbf{Deep Learning-based Countermeasure.}
Recent defenses apply deep learning to generate adversarial perturbations against traffic classifiers. GAN-based methods, such as \cite{li2019dynamic, sheffey2019improving}, craft indistinguishable flow features, though translating these into valid network traffic remains challenging due to protocol constraints. To overcome this, iPET \cite{shenoi2023ipet} and NIDSGAN \cite{zolbayar2022generating} directly perturb traffic at the packet level. Other approaches, including BAP \cite{nasr2021defeating}, insert dummy packets to obfuscate flow patterns \cite{rahman2020mockingbird, wang2017walkie, ling2022towards, juarez2015wtf}. However, these methods remain vulnerable to pre-trained classifiers. ET-BERT \cite{lin2022bert} achieves over 90\% accuracy across diverse traffic types, including encrypted traffic, rendering existing adversarial perturbations ineffective. To overcome this, we propose a novel perturbation method to evade ET-BERT’s traffic classification.

\section{Conclusion}\label{sec:Conclusion}

To defend against state-of-the-art traffic classification models, particularly transformer-based architectures, we begin by analyzing the intrinsic vulnerabilities of existing defense strategies. Building on these insights, we introduce an effective pre-padding strategy that modifies traffic semantics to mislead transformer-based classifiers into generating incorrect predictions. To further strengthen the adversarial effect, we incorporate reinforcement learning to optimize the perturbation process. Comprehensive experiments are conducted to evaluate the effectiveness of our proposed method, \ourNoSpace, across a range of scenarios, including white-box and black-box defenses, transferability evaluations, ablation studies, and parameter sensitivity analyses. The results demonstrate that our approach not only achieves robust defense performance across three real-world datasets but also exhibits strong practicality for deployment in operational network environments.

\bibliographystyle{plain}
\bibliography{ref}

\clearpage
\appendix

\begin{figure}[!t]
	\centering
	\includegraphics[scale=0.31]{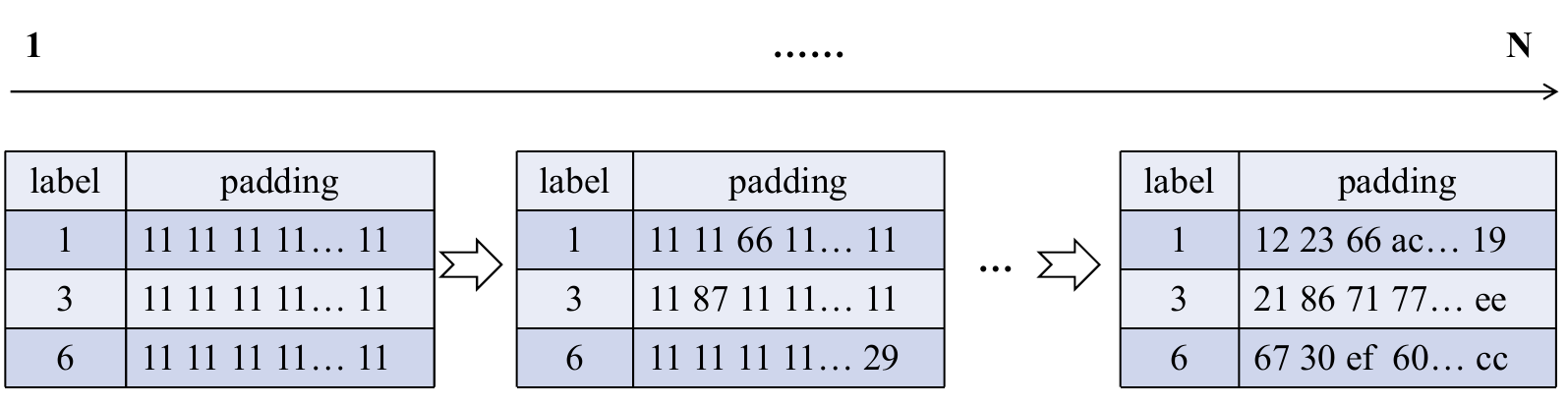}
	\caption{Evolution pattern of adversarial byte sequences with increasing temperature coefficient.}
	\label{fig:appendixtempetaure}
\end{figure}

\subsection{The Influence of Temperature Coefficient On The Action}\label{appendix:tempetaure}

As illustrated in Figure~\ref{fig:appendixtempetaure}, increasing the temperature coefficient results in a more uniform and less deterministic action distribution. Conversely, lower temperature values sharpen the distribution, accentuating the differences among candidate actions and leading to more confident and targeted perturbations generated by the model. However, such sharp perturbations are more likely to be perceived as abnormal by human observers, potentially increasing the risk of detection.

\end{document}